\documentclass[english,twocolumn]{revtex4-2}
\usepackage[T1]{fontenc}
\usepackage[utf8]{inputenc}
\usepackage{babel}
\usepackage{bm}
\usepackage{amsbsy}
\usepackage{amssymb}
\usepackage{graphicx}
\usepackage{esint}
\usepackage[pdfusetitle,
 bookmarks=true,bookmarksnumbered=false,bookmarksopen=false,
 breaklinks=false,pdfborder={0 0 1},backref=false,colorlinks=false]
 {hyperref}



\begin{document}
\title{Magnetization, excitations, and microwave power absorption in transition-metal/rare-earth
ferrimagnets with disorder}
\author{Dmitry A. Garanin and Eugene M. Chudnovsky}
\affiliation{Physics Department, Herbert H. Lehman College and Graduate School,
The City University of New York, 250 Bedford Park Boulevard West,
Bronx, New York 10468-1589, USA }
\date{\today}
\begin{abstract}
Efficient numerical routines are developed for numerical studies of
the dependence of the equilibrium magnetic states, excitations, and
microwave power absorption on temperature and composition in transition-metal/rare-earth
ferrimagnets, including the reversal of the Néel vector occurring
on both temperature and the concentration of the rare-earth atoms.
It results in a drastic change in the behavior at the magnetization
and angular-momentum compensation points. Dominant uniform oscillation
modes are obtained numerically by computing the magnetization correlation
function. They are compared with the analytical solution, which is
analyzed in detail. The fluctuation-dissipation theorem is used to
compute the frequency dependence of the absorbed microwave power.
A good agreement with analytical results is demonstrated. Disorder
caused by random positions of rare-earth atoms in a diluted RE system
leads to multiple localized modes that converge into broad absorption
maxima as the size of the system increases. The power absorption integrated
over frequency exhibits a minimum at the compensation point.
\end{abstract}
\maketitle

\section{Introduction}

Ferrimagnets are a class of magnetic materials formed by the antiferromagnetic
exchange interaction between two or more non-equivalent ferromagnetic
atomic sublattices \citep{Neel,Wolf}. A nonzero magnetization in
the absence of an external magnetic field makes ferrimagnets different
from antiferromagnets, which consist of identical ferromagnetic sublattices
with opposite magnetization. Antiferromagnets have long been considered
potential candidates for fast information technology applications
due to their fast spin dynamics arising from the compensation of the
total angular momentum associated with the spins of the sublattices.
This already becomes apparent from the fact that in antiferromagnets
the frequency of the uniform magnetic resonance is of order $\sqrt{DJ}/\hbar$,
while in ferromagnets it is of order $D/\hbar$, where $D$ is a constant
of the magnetic anisotropy which is relativistically small compared
to a much greater exchange interaction $J$ between ferromagnetic
sublattices, formed by the Coulomb forces \citep{Lectures}. However,
a compensated magnetic moment in antiferromagnets makes it difficult
to utilize their fast dynamics for application in devices.

In ferrimagnets, there are at least two different sublattices, usually
a transition-metal (TM) sublattice and a rare earth (RE) sublattice.
Correspondingly, there are two excitation modes, as in antiferromagnets.
However, as the sublattices are non-equivalent, there is a net magnetization.
If the difference between the sublattice angular momenta decreases
toward the compensation point on the RE concentration or on temperature,
the low-frequency mode stiffens while the high-frequency mode softens,
and these modes cross around the compensation point. In a certain
region around this point, there is a fast dynamics, as in antiferromagnets,
while the total magnetization is nonzero. As the gyromagnetic ratios
of the TM and RE spins are different, there are two different compensation
points: one for the angular momentum and one for the magnetic moment.
This is beneficial for applications, see e.g., Refs.\ {[}\onlinecite{Binder-PRB2006,Arena-PRApplied2017,Siddiqui-PRB2018,Ivanov-review,Bonfiglio-PRB2019,Kim-Nature2020,DavydovaJOP2020,Yurlov-PRB2021,Joo-materials2021,Chanda-PRB2021,Kim-NatMat2022,Guo-PRB2022,Zhang-PRB2022,Chen-Materials2025,Moreno-PRB2025,Ciccarelli-AP2025}{]},
because, unlike in antiferromagnets, it provides significant magnetization
in the region of compensation of the angular momentum.

Spin waves in ferromagnets were introduced by Felix Bloch \citep{Bloch}.
The discussion of uniform oscillations of the magnetization induced
by the ac magnetic field in the microwave frequency range -- the
Ferromagnetic Resonance (FMR) goes back to Griffiths \citep{Griffiths},
Kittel \citep{Kittel}, and Walker \citep{Walker-PRB1957}. Spin waves
in both ferro- and antiferromagnets were studied in great detail in
the book of Akhiezer, Bar'yakhtar, and Peletminskii \citep{ABP-SpinWaves},
see also Ref.\ {[}\onlinecite{Lectures}{]}. Early on, a detailed
computation of the frequencies of the magnetic resonance in ferrimagnets
appeared in the work of Wangsness \citep{Wangsness-PR1953} and was
discussed in relation to the magnetic resonance in rare-earth garnets
by Kittel \citep{Kittel-PR1959}. It was later re-derived and generalized
for spin waves with a finite wave vector by solving the Landau-Lifshitz
equation for classical spins and by diagonalizing quantum spin Hamiltonians:
see, e.g., Refs. {[}\onlinecite{Lin-PRB1988,Zhang-JPhys1997,Karchev-JPhys2008,Okuno-APLExpress2019,Haltz-PRB2022,Sanchez-PRB2025}{]}.
Numerous experiments, see, e.g., Refs.\ {[}\onlinecite{Pardavi-JMM2000,Binder-PRB2006,Stanciu-PRB2006,Okuno-APLExpress2019,Kim-Nature2020}{]},
confirmed theoretical expectations regarding hardening of the magnetic
resonance, and spin waves in general, on approaching the angular momentum
compensation point in ferrimagnets.

Compensation of the angular momentum in ferrimagnets also leads to
the increase of the domain-wall velocities, see, e.g., the experimental
work \citep{Siddiqui-PRB2018}. In Ref. \citep{Garanin1992}, nonlinear
mobility of the domain wall was calculated with the account of the
thermal disordering of the RE subsystem with the help of the Landau-Lifshitz-Bloch
(LLB) equation \citep{Garanin1998}. In Ref. \citep{Chubykalo2012},
ultrafast magnetization switching in ferrimagnets caused by a heat
pulse was studied using the LLB equation.

Motivation for our work is threefold. Firstly, while theoretical papers
on magnetic resonance in ferrimagnets have been abundant, studies
of their temperature dependence have been scarce. The latter is especially
important when exchange interaction between spins within the RE sublattice
is negligible, while the intersublattice coupling is weak relative
to the exchange in the TM sublattice. This is the case in widely studied
transition-metal/rare-earth ferrimagnetic compounds, such as FeGd,
CoCd, and FeCoGd alloys. At low temperature, RE spins are mostly aligned
opposite to the Fe and Co spins by the Fe-Gd and Co-Gd antiferromagnetic
exchange interaction. At elevated temperatures, they begin to fluctuate
widely and disorder stronger than the TM spins. This can result in
the inversion of the Néel vector (spin-flip), resulting in the total
magnetization being directed along the applied field.

Secondly, when addressing the problem of the magnetic resonance in
a diluted alloy, randomness in the positions of rare-earth atoms provides
another complication for computing the response of the system to a
microwave field. Due to the quenched randomness, the system exhibits
multiple spatially localized modes (see, e.g., Ref. \citep{GarChu2023})
that converge into a broadband absorption maximum on increasing the
size of the system.

Thirdly, we are interested in the absorption of microwave power by
a transition-metal/rare-earth ferrimagnetic alloy across the broad
range of temperatures and concentration of rare-earth, which may have
significant potential for applications. Notice that the association
of the accelerated dynamics with the hardening of one of the oscillation
modes on approaching the angular momentum compensation point has often
been attributed in literature to the divergence of the effective gyromagnetic
ratio. We find it useful to show that correct analytical formulas
and numerical results for magnetic resonances in ferrimagnets do not
support this conjecture.

To address the above problems numerically, we employ our own computational
methods for classical-spin systems implemented in Wolfram Mathematica.
These methods include: (i) fast energy minimization at $T=0$ combining
the sequential alignment of spins along their effective fields with
overrelaxation \citep{GarChuPro2013}; (ii) thermalized overrelaxation
\citep{GarChu2022} mandatory in systems with single-site anisotropy,
used in a combination with the Metropolis Monte Carlo at $T>0$; (iii)
adaptive Monte Carlo routine, the latest version of which can be found
in Ref. \citep{Garanin2025}; (iv) solution of the Landau-Lifshitz
equation of motion for conservative many-spin systems by high-order
solvers with the energy correction procedure needed to prevent energy
drift in long computations \citep{Garanin2021} that we modify here
for ferrimagnets. CoGd alloy has been chosen as a prototype. The energy
minimization on large spin lattices at $T=0$ and Monte Carlo simulations
at $T>0$ allow us to reveal the spin-flip transitions on temperature
and concentration of Gd. These procedures are combined with the study
of the dynamical evolution of the system needed to understand the
dependence of oscillation modes on temperature and rare-earth concentration.
We use the fluctuation-dissipation theorem to obtain the dependence
of the absorbed microwave power on the frequency of the ac field for
different rare-earth concentrations and different temperatures.

As our microscopic approach gives access to the intrinsic decoherence
and dissipation processes in the system due to the static randomness
and nonlinear interactions of spin excitations at elevated temperatures,
we do not include the phenomenological damping in the Landau-Lifshitz
equations in studying the dynamical evolution.

The article is organized as follows. The spin Hamiltonian and the
model are discussed in Section \ref{Sec_The-model}. Section \ref{Sec_Excitation-modes}
presents an analytical derivation of the frequencies of the magnetic
resonance for the model in hand, their analysis far from and close
to the compensation points, and computation of the magnetic fields
corresponding to the re-orientation of the magnetizations of sublattices.
Section \ref{Sec_Numerical-methods} contains details of the numerical
methods, such as the energy minimization, Metropolis Monte Carlo,
and the solution of the Landau-Lifshitz equation, as well as the energy
correction algorithms, definition of the spin temperature in a ferrimagnet,
and specifics of the parallelized computation with Wolfram Mathematica.
Sections \ref{Sec_Compensation} and \ref{Sec_Inversion-Neel} are
devoted to the compensation of the angular momentum and magnetic moment,
as well as to the behavior of the magnetization and spin-flip transitions
on temperature and concentration of rare-earth atoms. Uniform excitation
modes are computed from the magnetization correlation function and
compared with analytical results in Section \ref{Sec_Uniform-excitation-modes}.
Microwave absorption spectra are computed in Section \ref{Sec_The-absorption-spectrum}.
Numerical results on the integral power absorption, corresponding
to the integral over all frequencies, are presented in Section \ref{Sec_The-integral-absorption}.

\section{The model}

\label{Sec_The-model}

Consider the simplest two-dimensional lattice model of a ferrimagnet,
which consists of the \textquotedbl parent\textquotedbl{} sublattice
built of transition metal (TM) atoms, for instance, Fe or Co, exhibiting
all ferromagnetic features, and a Rare-Earth (RE) sublattice of loose
spins coupled antiferromagnetically to the parent spins. The Hamiltonian
reads
\begin{eqnarray}
\mathcal{H} & = & -\frac{1}{2}\sum_{ij}J_{ij}\mathbf{S}_{i}\cdot\mathbf{S}_{j}-\mathbf{H}\cdot\sum_{i}\left(g\mu_{B}\mathbf{S}_{i}+g'\mu_{B}\boldsymbol{\sigma}_{i}\right)\nonumber \\
 &  & +J'\sum_{i}\mathbf{S}_{i}\cdot\boldsymbol{\sigma}_{i}+\mathcal{H}_{\mathrm{other}},\label{Ham_ferrite}
\end{eqnarray}
where $\mathbf{S}_{i}$ are parent spins interacting with each other
by the nearest-neighbor exchange with the exchange constant $J>0$
in a square lattice, $\mathbf{H}$ is the applied magnetic field,
$\boldsymbol{\sigma}_{i}$ are RE spins coupled to the TM spins by
the exchange constant $J'>0$, while $g$ and $g'$ are g-factors
for both kinds of spins. $\mathcal{H}_{\mathrm{other}}$ includes
all other interactions in the parent sublattice such as anisotropy,
Dzyaloshinskii-Moriya interaction (DMI), etc.:
\begin{eqnarray}
\mathcal{H}_{\mathrm{A}} & = & -\frac{D}{2}\sum_{i}S_{i,z}^{2},\label{Ham_A}\\
\mathcal{H}_{\mathrm{DMI}} & = & A\sum_{i}\left[\left(\mathbf{S}_{i}\times\mathbf{S}_{i+\delta_{x}}\right)_{x}+\left(\mathbf{S}_{i}\times\mathbf{S}_{i+\delta_{y}}\right)_{y}\right],\label{H_DMI}
\end{eqnarray}
where $\mathcal{H}_{\mathrm{DMI}}$ is of the Bloch-type DMI with
$i+\delta_{x,y}$ being the neighboring sites in the positive $x$
and $y$ directions. This kind of DMI favors skyrmions with the counterclockwise
orientation of the in-plane spin components for $A>0$. Here we consider
the DMI originating from interactions in the bulk of the film and
neglect the interfacial DMI that depends on the composition and film
thickness. For the TM/RE systems we are studying, the bulk DMI may
be dominant \citep{Berges2022}. The spin lengths are considered to
be different, $S<\Sigma$, which results in different angular momenta
of the spins, $\hbar S$ and $\hbar\Sigma$. The magnetic moments
of the spins are given by $\mu=g\mu_{B}S$ and $\mu'=g'\mu_{B}\Sigma$.
For Co $S=3/2$ and $g=2.2,$whereas for Gd $\Sigma=7/2$ and $g=2$.
Here, spins are considered as classical vectors of lengths $S$ and
$\Sigma$.

The continuous version of this lattice model has the energy density
given by
\begin{equation}
\epsilon=\epsilon_{\mathrm{ex}}+\epsilon_{Z}+\epsilon_{A}+\epsilon_{\mathrm{DMI}}+\ldots,
\end{equation}
where
\begin{eqnarray}
\epsilon_{\mathrm{ex}} & = & \frac{1}{a^{d-2}}\frac{J}{2}\bm{\nabla}s_{\alpha}\cdot\bm{\nabla}s_{\alpha}+\frac{1}{a^{d}}J'\mathbf{S}\cdot\boldsymbol{\sigma}\nonumber \\
\epsilon_{Z} & = & -\frac{1}{a^{d}}\mathbf{H}\cdot\left(g\mu_{B}\mathbf{S}+g'\mu_{B}\boldsymbol{\sigma}\right)\nonumber \\
\epsilon_{A} & = & -\frac{1}{a^{d}}\frac{D}{2}S_{z}^{2}\nonumber \\
\epsilon_{\mathrm{DMI}} & = & -\frac{1}{a^{d-1}}A\mathbf{S}\cdot\left(\nabla\times\mathbf{S}\right),\label{Energy_density}
\end{eqnarray}
where $a$ is the lattice spacing in a hypercubic lattice in the dimension
$d$ (here $d=2$), the summation over repeated indices is implied
in $\epsilon_{\mathrm{ex}}$, and in $\epsilon_{\mathrm{DMI}}$ the
derivatives over $z$ should be discarded. Note that the coefficients
in different terms in the energy density have different units, which
makes it difficult to compare their strength. If the lattice structure
is complicated and the parameters are extracted from macroscopic measurements,
the continuous form of the energy is considered as primary, and the
lattice version above as its discretization. Lattice computations
on realistic (nonhypercubic) lattices are rare. In most cases, the
hypercubic lattice model with a microscopic value of $a$ provides
a reasonable description of magnetic properties at the atomic scale,
including the effects of thermal disordering. To the contrary, micromagnetic
computations use hypercubic discretizations with much larger (mesoscopic
to macroscopic) values of the discretization parameter and cannot
reliably describe thermal effects, which require an atomistic approach.
Equations (\ref{Energy_density}) can be also expressed in terms of
the TM and RE spin densities $\boldsymbol{\rho}_{S}=\mathbf{S}/a^{d}$
and $\boldsymbol{\rho}_{\Sigma}=\boldsymbol{\sigma}/a^{d}$. We will
not use the continuous model here and show it only for a comparison
with other publications.

The equations of motion for the lattice spins have the form
\begin{equation}
\mathbf{\dot{S}}_{i}=\frac{1}{\hbar}\left[\mathbf{S}_{i}\times\mathbf{H}_{\mathrm{eff},i}\right],\qquad\dot{\boldsymbol{\sigma}}_{i}=\frac{1}{\hbar}\left[\boldsymbol{\sigma}_{i}\times\mathbf{H}'_{\mathrm{eff},i}\right]\label{Larmor}
\end{equation}
with the effective fields given by
\begin{eqnarray}
\mathbf{H}_{\mathrm{eff},i} & \equiv & -\frac{\partial\mathcal{H}}{\partial\mathbf{S}_{i}}=\sum_{j}J_{ij}\mathbf{S}_{j}+g\mu_{B}\mathbf{H}+DS_{iz}\mathbf{e}_{z}-J'\boldsymbol{\sigma}_{i}\nonumber \\
\mathbf{H}'_{\mathrm{eff},i} & = & g'\mu_{B}\mathbf{H}+D_{\Sigma}\sigma_{iz}\mathbf{e}_{z}-J'\mathbf{S}_{i}.\label{Heff}
\end{eqnarray}
Here, the uniaxial anisotropy in the RE sublattice was added for generality.
As $D_{\Sigma}$ is typically very small, it will be discarded later.

The model can be generalized by the dilution of RE atoms. For this,
one can in the Hamiltonian replace $\boldsymbol{\sigma}_{i}\rightarrow p_{i}\boldsymbol{\sigma}_{i}$,
where $p_{i}$ are random occupation numbers, $p_{i}=0,1$. The RE
concentration is defined by $c=\left\langle p_{i}\right\rangle $
and it is changing between 0 and 1. Diluting the RE system allows
one to create the compensation of the angular momentum at $c\Sigma=S$,
as well as the compensation of the magnetic moment at $c\Sigma=\left(g/g'\right)S$.
Diluting the TM subsystem is also possible, but it is not done here,
because the model formulated above already describes the compensation
transition on $c$. Keeping in mind thin ferrimagnet films, here we
consider the two-dimensional lattice model.

\section{Excitation modes in the uniform state}

\label{Sec_Excitation-modes}

The spectrum of excitation modes in a ferrimagnet can be calculated
by linearizing the equations of motion in small deviations from the
ground state and making the Fourier transformation. We consider the
anticollinear ground state aligned along the $z$ axis, $S_{iz}=S$
and $\sigma_{iz}=-\Sigma$ in the model with $\mathbf{H}=H\mathbf{e}_{z}$.
The calculation shown in the Appendix results in the secular equation
for the energy spectrum
\begin{eqnarray}
\left(\varepsilon_{TM}-\varepsilon\right)\left(\varepsilon_{RE}-\varepsilon\right)+cS\Sigma J'{}^{2} & = & 0,\label{Secular_equation}
\end{eqnarray}
where
\begin{eqnarray}
\varepsilon_{TM} & \equiv & S\left(J_{0}-J_{\mathbf{k}}+D\right)+g\mu_{B}H+c\Sigma J'\\
\varepsilon_{RE} & \equiv & -\Sigma D_{\Sigma}+g'\mu_{B}H-J'S.
\end{eqnarray}
The solution of this quadratic equation for $\varepsilon$ is
\begin{equation}
\varepsilon_{\pm}=\frac{1}{2}\left[\varepsilon_{TM}+\varepsilon_{RE}\pm\sqrt{\left(\varepsilon_{TM}-\varepsilon_{RE}\right)^{2}-4c\Sigma SJ'{}^{2}}\right].\label{epsilon_pm-general}
\end{equation}
The sign of $\varepsilon_{\pm}$ is related to the direction of spin
precession. The DMI does not make a contribution into the modes' frequencies
because for the collinear ground state there are no linear deviation
terms due to the DMI.

In the quasi-symmetric case $c=1$, $\Sigma=S$, and $D_{\Sigma}=D$,
in zero field, the result for the uniform mode, $k=0$, simplifies
to
\begin{equation}
\varepsilon_{\pm}=\pm S\sqrt{D\left(2J'+D\right)},\label{eps_pm_AFM}
\end{equation}
same as the antiferromagnetic degenerate modes. Below, we consider
the case $D_{\Sigma}=0$.

For the $H=0$ uniform modes, if the condition $SD+\left(c\Sigma-S\right)J'=0$,
i.e.,
\begin{equation}
c=\frac{S}{\Sigma}\left(1-\frac{D}{J'}\right),
\end{equation}
is fulfilled, the spectrum becomes
\begin{equation}
\varepsilon_{\pm}=\pm S\sqrt{DJ'}.
\end{equation}
Again, the frequencies of the two modes coincide up to their sign.

At the angular-momentum compensation point, $c=S/\Sigma$, one has
\begin{equation}
\varepsilon_{\pm}=\frac{S}{2}\left(D\pm\sqrt{D\left(4J'+D\right)}\right).
\end{equation}
Typically $J'\gg1D$, so that the second term under the square root
can be dropped. One can see that the mode splitting at compensation
is $\varepsilon_{+}-\left|\varepsilon_{-}\right|=SD.$ For $SJ=1,$$SJ'=0.2$
and $SD=0.03$ one has $\varepsilon_{+}=0.0925$ and $\varepsilon_{-}=-0.0625$.

If the intersublattice coupling is strong and the system is far from
the compensation point, one of the modes $\varepsilon_{\pm}$ has
a low frequency (LF) and another one has a high frequency (HF). In
this case, one can expand Eq. (9) (for $k=0$ and $H=0$) in powers
of $D/J'$ to obtain, up to the second order of the perturbation theory,
\begin{equation}
\varepsilon_{\mathrm{HF}}=\left(c\Sigma-S\right)J'+\frac{c\Sigma SD}{c\Sigma-S}
\end{equation}
and
\begin{equation}
\varepsilon_{\mathrm{LF}}=\frac{S^{2}D}{S-c\Sigma}-\frac{c\Sigma S^{3}D^{2}}{(S-c\Sigma)^{3}J'}.
\end{equation}
For $S-c\Sigma>0$ one has $\varepsilon_{\mathrm{HF}}=\varepsilon_{-}$
and $\varepsilon_{\mathrm{LF}}=\varepsilon_{+}$, while for $S-c\Sigma<0$
one has $\varepsilon_{\mathrm{HF}}=\varepsilon_{+}$ and $\varepsilon_{\mathrm{LF}}=\varepsilon_{-}$.
Up to the direction of spin precession, the sign of $\varepsilon$
is irrelevant. Toward the compensation point, the HF mode softens
while the LF mode stiffens, but then both modes meet, and the approximation
above becomes invalid. In this region, the frequencies of both modes,
$\varepsilon\sim S\sqrt{DJ'}$, are defined by the exchange-enhanced
anisotropy. The first term of the formula above with the the inhomogeneous
exchange and the magnetic field added reads
\begin{equation}
\varepsilon_{\mathrm{LF}}=\frac{S^{2}\left(J_{0}-J_{\mathbf{k}}+D\right)+\left(gS-cg'\Sigma\right)\mu_{B}H}{S-c\Sigma}.
\end{equation}
This is quasi-ferromagnetic mode.

A sufficiently strong magnetic field causes ground-state transitions
driven by the instability of the collinear state. At these transitions,
the frequency of one of the modes should go to zero at $k=0$. From
the first line in Eq. (\ref{epsilon_pm-general}), one can see that
$\varepsilon_{-}=0$ if $\varepsilon_{TM}\varepsilon_{RE}+c\Sigma SJ'{}^{2}=0$.
From this condition follows
\begin{eqnarray}
H_{\pm} & = & \frac{1}{2g\mu_{B}g'\mu_{B}}\left\{ -g'\mu_{B}SD+\mu J'\right.\nonumber \\
 &  & \left.\pm\sqrt{\left[g'\mu_{B}SD-\mu J'\right]^{2}+4gg'\mu_{B}^{2}S^{2}DJ'}\right\} ,\label{Hpm_general}
\end{eqnarray}
where
\begin{equation}
\mu\equiv g\mu_{B}S-g'\mu_{B}c\Sigma\label{mu_def}
\end{equation}
is the magnetic moment. In real cases $D\ll J'$, so that not close
to the magnetic moment compensation point, $\Delta\mu=0$, one can
expand this expression to obtain
\begin{equation}
H_{+}\cong\frac{S^{2}D}{\mu},\qquad H_{-}=\frac{\mu J'}{gg'\mu_{B}^{2}}.\label{Hpm_far_from_compensation}
\end{equation}
The small $H_{+}$ corresponds to the spin-flip of the sublattices
with respect to the anisotropy axis (the inversion of the Néel vector).
The large $H_{-}$ corresponds to the spin-flop of the sublattices
with respect to each other. Note that we investigated the state with
$S_{iz}=S$ and $\sigma_{iz}=-\Sigma$. For the flipped state with
$S_{iz}=-S$ and $\sigma_{iz}=\Sigma$, the signs in front of $S$
and $\Sigma$ in all formulas above should be inverted. Thus for the
spin-flop instability field one has a general formula
\begin{equation}
H_{-}=\pm\frac{\left|\mu\right|J'}{gg'\mu_{B}^{2}}.\label{H_spin-flop}
\end{equation}
If a magnetic field $H$ is applied, then the spin-flop phase exists
in the range
\begin{equation}
\left|\mu\right|\leq gg'\mu_{B}^{2}H/J'\label{mu_spin-flop_region}
\end{equation}
around the magnetic-moment compensation point $\mu=0$. In the absence
of the anisotropy, there is a spin-flip transition in any weak field
$H$ on $\mu$: the spins reorient so that the total magnetic moment
is collinear with $\mathbf{H}$. For a finite $H,$the spin-flip transition
goes via the intermediate spin-flop phase. The latter is energetically
favorable, because canting of the sublattices in the presence of the
magnetic field leads to decreasing of the energy.

In the presence of the anisotropy, at magnetic-moment (MM) compensation
point from Eq. (\ref{Hpm_general}) one obtains
\begin{equation}
H_{\pm}\cong\pm\frac{S\sqrt{DJ'}}{\mu_{B}\sqrt{gg'}}.\label{Hpm_at_compensation}
\end{equation}
Here, the anisotropy is exchange-enhanced.

Analytical consideration of the model with the magnetic field applied
non-collinear with the anisotropy axis and of the spin-flop phase
requires more efforts. The results of this section can be extended
to non-zero temperatures with the help of the mean-field approximation,
as was done, e.g., in Ref. \citep{Haltz-PRB2022}.

\section{Numerical methods}

\label{Sec_Numerical-methods}

Numerical calculations were performed in the lattices of sizes $N_{x}\times N_{y}$
with periodic boundary conditions, and on each lattice site there
were two spins: a TM spin and a RE spin. The total number of sites
is $\mathcal{N}=N_{x}N_{y}$, while the total number of spins is $2\mathcal{N}$.

We perform three types of computations: i) Energy minimization at
$T=0$ to study the spin-flip transition in a magnetic field in the
absence of the anisotropy on the RE concentration $c$; ii) Monte
Carlo simulations at $T>0$ to study this transition on temperature;
iii) Dynamical evolution according to Eq. (\ref{Larmor}) to elucidate
the concentration and temperature dependences of the uniform modes.

The energy minimization combines aligning the spin $\mathbf{s}_{i}$
with its effective field $\mathbf{H}_{\mathrm{eff},i}$ with the probability
$\alpha$ and flipping the spin around the effective field, $\mathbf{S}_{i}\Rightarrow2\left(\mathbf{S}_{i}\cdot\mathbf{H}_{\mathrm{eff},i}\right)\mathbf{H}_{\mathrm{eff},i}/H_{\mathrm{eff},i}^{2}-\mathbf{S}_{i}$
with the probability $1-\alpha$ (the so-called overrelaxation) \citep{GarChuPro2013};
similar for the RE spins. The algorithm uses vectorized updates of
columns of spins in checkered sublattices, which allows parallelization
of the computation.

The Metropolis Monte Carlo routine for classical spin vectors includes
adding a random vector to a spin and normalizing the result to obtain
the trial configuration. After that, the energy change $\Delta E$
is computed and the new spin value is accepted if $\exp\left(-\Delta E/T\right)>\mathrm{rand}$,
where rand is a random number in the interval $\left(0,1\right)$.
If $\Delta E<0$, the trial is automatically accepted. Also here,
updating spins is done in the vectorized and parallelized form for
checkered sublattices. Combining Monte Carlo updates with overrelaxation
greatly speeds up the thermalization. For systems with single-site
anisotropy, overrelaxation does not conserve the energy, leading to
skewed results. In this case, one needs to perform the thermalized
overrelaxation \citep{GarChu2022}. The number of Monte Carlo steps
needed to thermalize the system greatly increases near phase transition
points. For this reason, to obtain reliable results across wide temperature
ranges, we used the adaptive Monte Carlo routine, the latest version
of which is described in Ref. \citep{Garanin2025}.

To compute the dynamical evolution of the system (at arbitrary temperatures),
we employed the fifth-order Butcher's Runge-Kutta ordinary differential
equation (ODE) solver, RK5, which makes six function evaluations per
integration step (see, e.g., the Appendix in Ref. \citep{Garanin2017}).
It is much more precise than the classical RK4 solver. Precision in
dynamical computation is important, since numerical errors accumulate
over a large evolution period, leading to the energy drift in conservative
systems under consideration. Whatever the accuracy of the ODE solver,
one needs to do the energy correction \citep{Garanin2021} from time
to time. Correcting the energy causes small deviations from the equilibrium
state. Nevertheless, if the system is ergodic, that is, it dynamically
thermalizes, the equilibrium state is restored and the method works
very well.

In this particular model of ferrimagnets, there is a problem with
correcting the energy. If the energy of the whole system is being
corrected, the energy migrates from one sublattice to the other, violating
thermal equilibrium. This apparently happens because RE spins do not
interact with each other and thus they do not form an ergodic large
subsystem that could dynamically thermalize. Because of this, one
has to do the detailed energy correction, that is, correct the energy
of the TM sublattice (without the intersublattice interaction) and
then correct that of the RE sublattice (including the intersublattice
interaction). Both target energies in the correction procedure are
equilibrium energies of the subsystems in the initial state, computed
by Monte Carlo. Since the energies of the subsystems fluctuate, one
has to perform the energy correction not too frequently, so that the
energy drift is corrected and not the energy fluctuations. This is
a difference from correcting the whole energy of a conservative system,
which can be done at any time.

To control the quality of the dynamical solution, we computed the
spin temperature (see Ref. \citep{Garanin2021} and references therein),
which for ferrimagnets with diluted RE subsystem has the form
\begin{equation}
T_{S}=\frac{\sum_{i}\left[\left(\mathbf{S}_{i}\times\mathbf{H}_{\mathrm{eff},i}\right)^{2}+\left(p_{i}\boldsymbol{\sigma}_{i}\times\mathbf{H'}_{\mathrm{eff},i}\right)^{2}\right]}{\sum_{i}\left(\mathbf{S}_{i}\cdot\mathbf{H}_{\mathrm{eff},i}+p_{i}\boldsymbol{\sigma}_{i}\cdot\mathbf{H'}_{\mathrm{eff},i}\right)+D\sum_{i}\left(S_{i,z}^{2}-S^{2}\right)}.\label{Ts}
\end{equation}
For the states thermalized by Monte Carlo at the set temperature $T$,
the value of $T_{S}$ is close to $T$ up to fluctuations that decrease
with the system's size. If the solution of dynamical equations is
accurate, $T_{S}$ remains close to $T$ at all times.

There is one more problem with dynamics for the systems we are studying.
The Hamiltonian containing the exchange interaction and uniaxial anisotropy
conserves the $z$ component of the total angular momentum. Integrals
of motion additional to the energy are detrimental to the ergodicity
of the system, because they do not dynamically relax to their equilibrium
values. What happens here is that energy corrections, performed from
time to time, change the $z$ component of the total angular momentum,
and these changes accumulate, causing a drift of this integral of
motion and violating thermal equilibrium. As a result, $T_{S}$ significantly
deviates from $T$. Fortunately, this problem can be solved by a trick.
By adding the DMI (which does not affect excitation modes in the uniform
state), one breaks the conservation of the $z$ component of the total
angular momentum and makes the system ergodic. With the DMI added,
this integral of motion dynamically relaxes back to its equilibrium
value and $T_{S}\cong T$ at all times. The value of the DMI should
be below the instability threshold of the uniform state, theoretically
\citep{Leonov}
\begin{equation}
\kappa\equiv\frac{\pi A}{\sqrt{8JD}}<1.\label{kappa_def}
\end{equation}
In our computations, we use $A/J=0.1$ and $D/J=0.03$, so that $\kappa=0.524$
and the uniform state is stable. Another but related virtue of the
DMI is that it provides an additional source for the relaxation processes
in the system. This makes the dynamics at elevated temperatures more
realistic, as in reality there are different kinds of nonlinear processes,
including ones that do not conserve the $z$ component of the total
spin.

Computations were performed in Wolfram Mathematica with core routines
vectorized and parallelized. For this, the spin tensor describing
the whole system was written as $s[[n_{x},\alpha,n_{y}]]$, where
$n_{x}=1,\ldots,N_{x},$$n_{y}=1,\ldots,N_{y}$, and $\alpha=1,\ldots,6$
are spin components ($\alpha=1,2,3$ for the TM spins and $\alpha=4,5,6$
for the RE spins). The operations were parallelized in $n_{x}$ and
vectorized in $n_{y}$ (operations done on the whole $n_{y}$ columns
without loops in $n_{y}$). As a result, the computations were rather
fast and memory-bound (the speed was limited by reading and writing
the memory, with a low processor utilization).

In the computations, we set, as usual, $\hbar=k_{B}=g\mu_{B}=1$,
as well as $SJ=1$. Throughout the paper, we use $S=3/2$, $\Sigma=7/2$,
$g'/g=2/2.2=0.909.$ The default value of the intersublattice coupling
is $J'/J=0.2$, although in some cases we use $J'/J=0.1$.

\section{Compensation of the angular momentum and magnetic moment}

\label{Sec_Compensation}

The first thing to investigate is the dependence of the magnetic state
on the RE concentration $c$ and the temperature $T$. Here we use
our energy-minimization routine at $T=0$ and Metropolis Monte Carlo
at $T>0$. We define the angular momentum per site as
\begin{equation}
\mathbf{m}=\frac{1}{\mathcal{N}}\sum_{i}\left(\mathbf{S}_{i}+\mathbf{\sigma}_{i}\right),\label{m_vec_def}
\end{equation}
and the normalized magnetic moment per site as
\begin{equation}
\boldsymbol{\mu}=\frac{1}{\mathcal{N}}\sum_{i}\left(\mathbf{S}_{i}+\frac{g'}{g}\mathbf{\sigma}_{i}\right).\label{mu_vec_def}
\end{equation}
If a magnetic field $\mathbf{H}$ is applied along the $z$ axis,
the system tends to lower its energy so that the total magnetic moment
is directed along $\mathbf{H}$. As the RE spin $\Sigma$ is larger
than the TM spin $S$, the total magnetic moment $\mu=g\mu_{B}S-g'\mu_{B}c\Sigma$
changes its sign on $c$ at the magnetic-moment (MM) compensation
point, and the spins should flip, inverting the Néel vector $\mathbf{l}=\mathbf{S}-c\boldsymbol{\Sigma}$.
In the presence of the uniaxial anisotropy, this transition is hampered
by the energy barrier between the stable and metastable states that
causes a hysteresis with the spin-flip field given by $H_{+}$ in
Eq. (\ref{Hpm_far_from_compensation}). For $H=0$ in the presence
of the uniaxial anisotropy, there is no spin-flip transition, and
the temperature dependences of the angular momentum and magnetic moment
are smooth. If the concentration of RE atoms is large enough, they
dominate magnetic properties at low temperatures. With increasing
the temperature, the RE subsystem disorders faster, than the TM subsystem,
which leads to the compensation of the angular momentum, $m_{z}=0$,
at the temperature $T_{A}$ and to the compensation of the magnetic
moment, $\mu_{z}=0$, at another temperature $T_{M}$. If $g'<g$,
then $T_{M}<T_{A}$. This behavior is shown for different concentrations
of the RE atoms, $c=1$ and $c=0.7$, in Fig. \ref{Fig_mz_mu_z}.
For the intersublattice coupling $J'/J=0.2,$the compensation points
correspond to rather high temperatures, which are comparable to the
magnetic phase-transition temperature, here $T_{C}/(SJ)\approx1$.
For $J'/J=0.1$, the RE subsystem disorders faster, and the compensation
occurs at lower temperatures. Compensation also occurs on changing
$c$.
\begin{center}
\begin{figure}
\begin{centering}
\includegraphics[width=8cm]{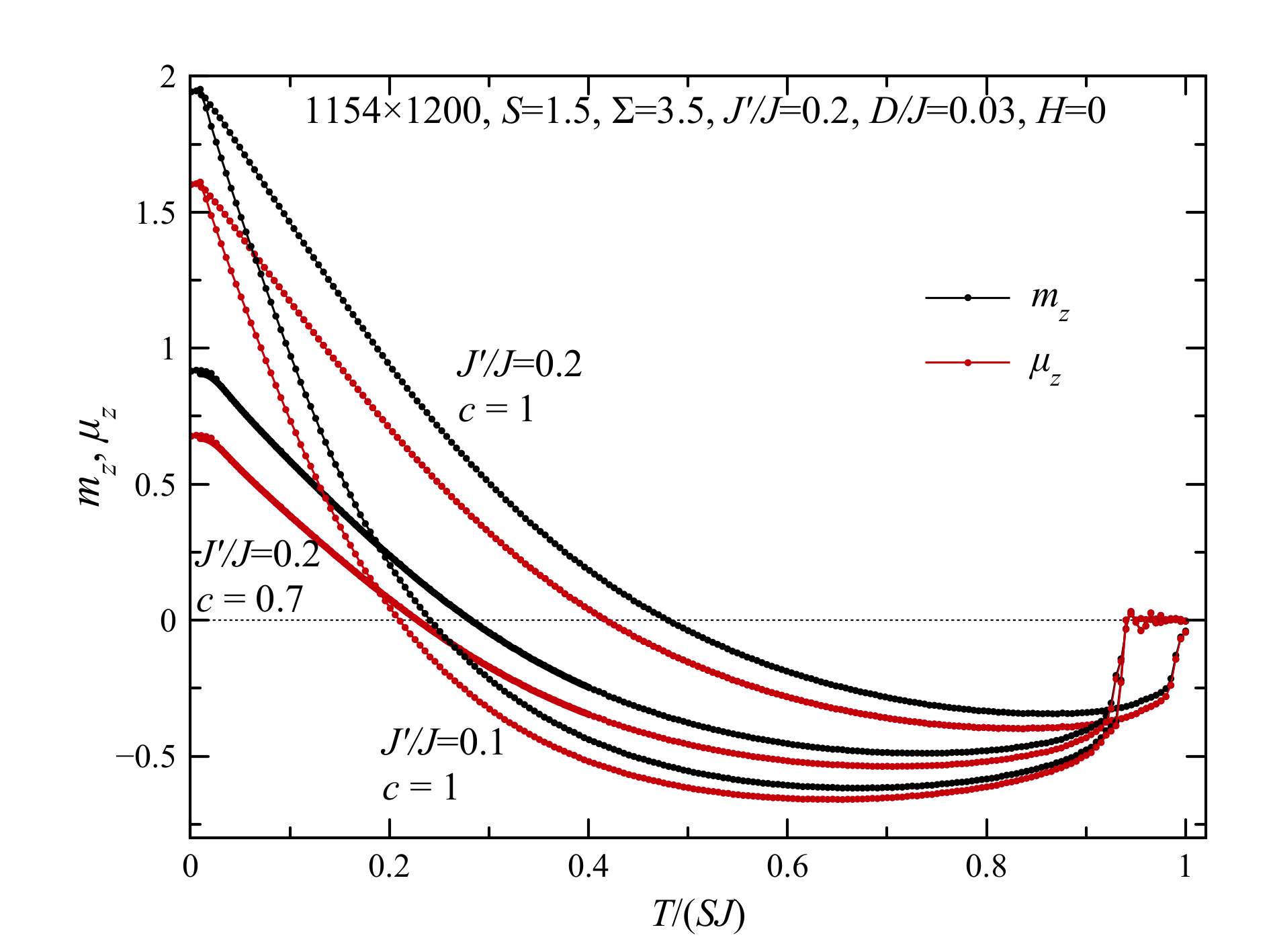}
\par\end{centering}
\caption{Total angular momentum ($m_{z}$) and total magnetic moment ($\mu_{z}$)
vs temperature, showing compensation of the angular momenta ($m_{z}=0$)
and magnetic moment ($\mu_{z}=0$) for different concentrations of
the RE atoms, $c=1$, $0.7$, and different intersublattice coupling,
$J'/J=0.1$, 0.2. Smaller intersublattice coupling $J'$ results in
a faster thermal disordering of the RE subsystem with increasing the
temperature. }\label{Fig_mz_mu_z}
\end{figure}
\par\end{center}

\section{Inversion of the Néel vector on the RE concentration and temperature}

\label{Sec_Inversion-Neel}

\begin{figure}
\begin{centering}
\includegraphics[width=8cm]{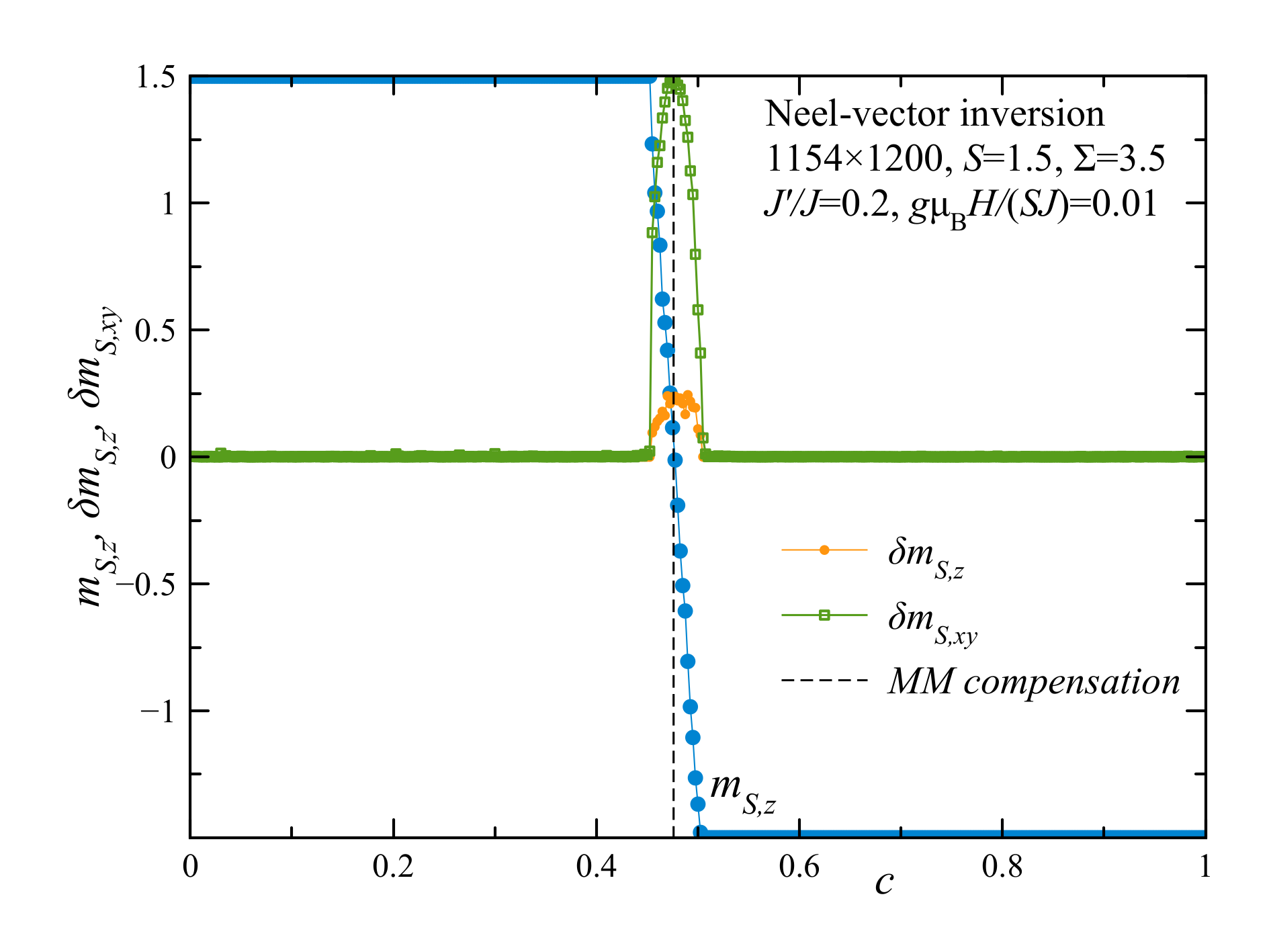}
\par\end{centering}
\begin{centering}
\includegraphics[width=8cm]{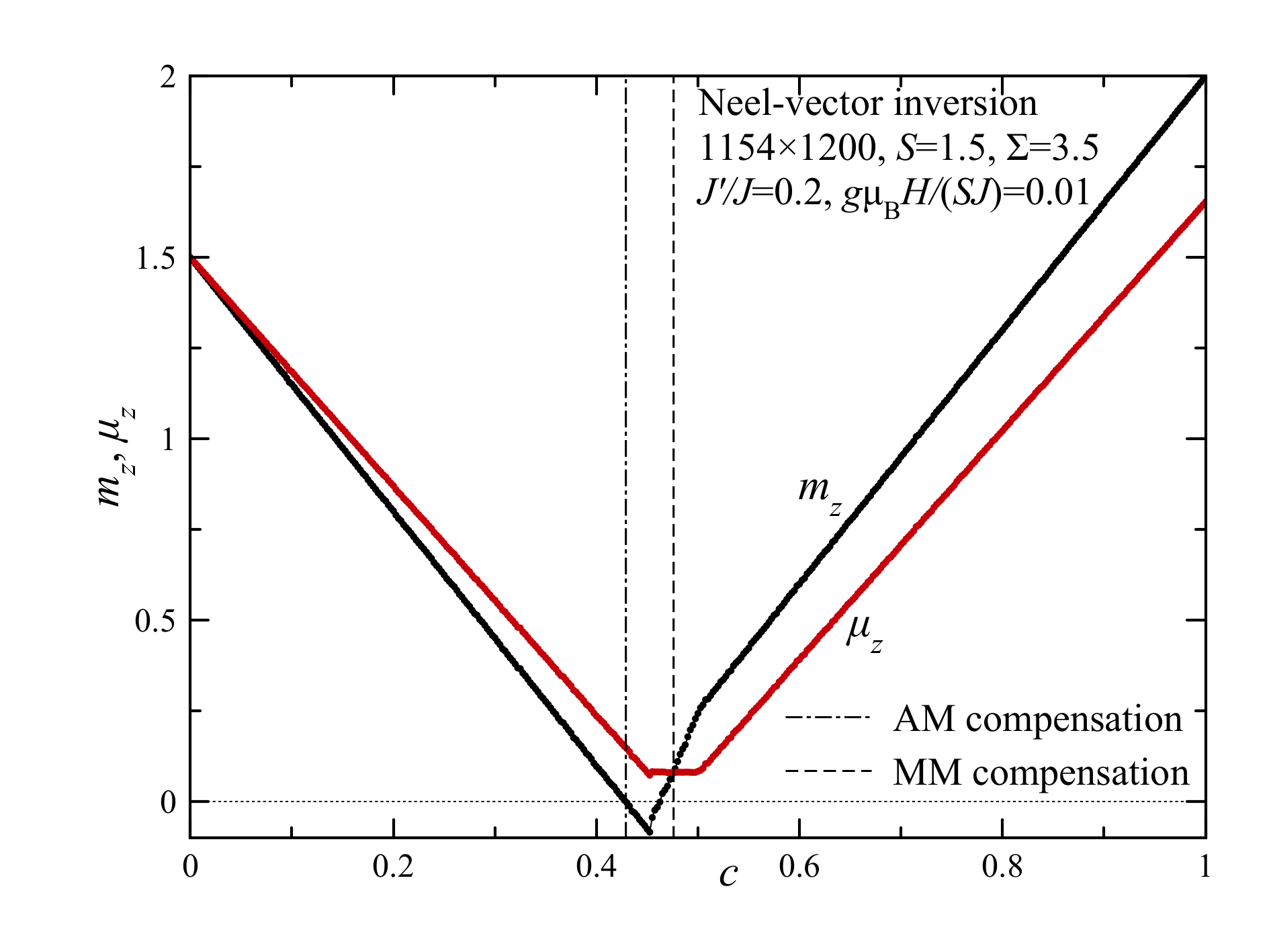}
\par\end{centering}
\caption{Spin-flip transition on the RE concentration $c$ via the intermediate
spin-flop phase, see Eq. (\ref{mu_spin-flop_region}). Top: Angular
momentum of the TM sublattice, $m_{z}$, and dispersion of the longitudinal
and transverse fluctuations. Bottom: Total angular momentum ($m_{z}$)
and total magnetic moment ($\mu_{z}$). }\label{Fig_Spin-flip-on_c}
\end{figure}

If $H$ exceeds $H_{+}$ given by Eq. (\ref{Hpm_at_compensation}),
the hysteresis disappears and the results become qualitatively similar
to those for $D=0$. Here we consider, for simplicity, the case of
a zero anisotropy, starting from the sweep over the RE concentration
$c$ at $T=0$. To find the ground state, we performed the energy
minimization for the systems with different RE $c$ starting from
the collinear initial state with the TM spins aligned along the magnetic
field $H/(SJ)=0.01$ and the RE spins aligned opposite to it. In fact,
as there are no energy barriers, the initial state can be arbitrary,
and the system comes to the same final state. The results for the
$z$ component of the average TM spin, $m_{S,z}=(1/\mathcal{N})\sum_{i}S_{i,z}$
and the dispersion of its longitudinal and transverse fluctuations
\begin{eqnarray}
\delta m_{S,z} & \equiv & \sqrt{\frac{1}{\mathcal{N}}\sum_{i}\left(S_{i,z}-m_{S,z}\right)^{2}}\nonumber \\
\delta m_{S,xy} & \equiv & \sqrt{\frac{1}{\mathcal{N}}\sum_{i}\left(S_{i,x}^{2}+S_{i,y}^{2}\right)}\label{deltamSxyz}
\end{eqnarray}
are shown in Fig. \ref{Fig_Spin-flip-on_c}(top). One can see that
the spin-flip transition occurs within a range of $c$, where $m_{S,z}$
changes continuously between $S$ and $-S$. The average RE spin's
$z$ projection changes in this region between $-\Sigma$ and $\Sigma$
(not shown). In the transient region, the dispersion $\delta m_{S,z}$
is significant. This is an apparent consequence of the fluctuations
of the local RE concentration. We have found that the transient region
broadens with increasing $H$, in accordance with Eq. (\ref{mu_spin-flop_region}).
This and the large value of $\delta m_{S,xy}$ suggest that the intermediate
phase is a spin-flop phase. The spin canting caused by $H$ leads
to the energy gain; this is why the region of the spin-flop phase
broadens with $H$.

\begin{figure}
\begin{centering}
\includegraphics[width=8cm]{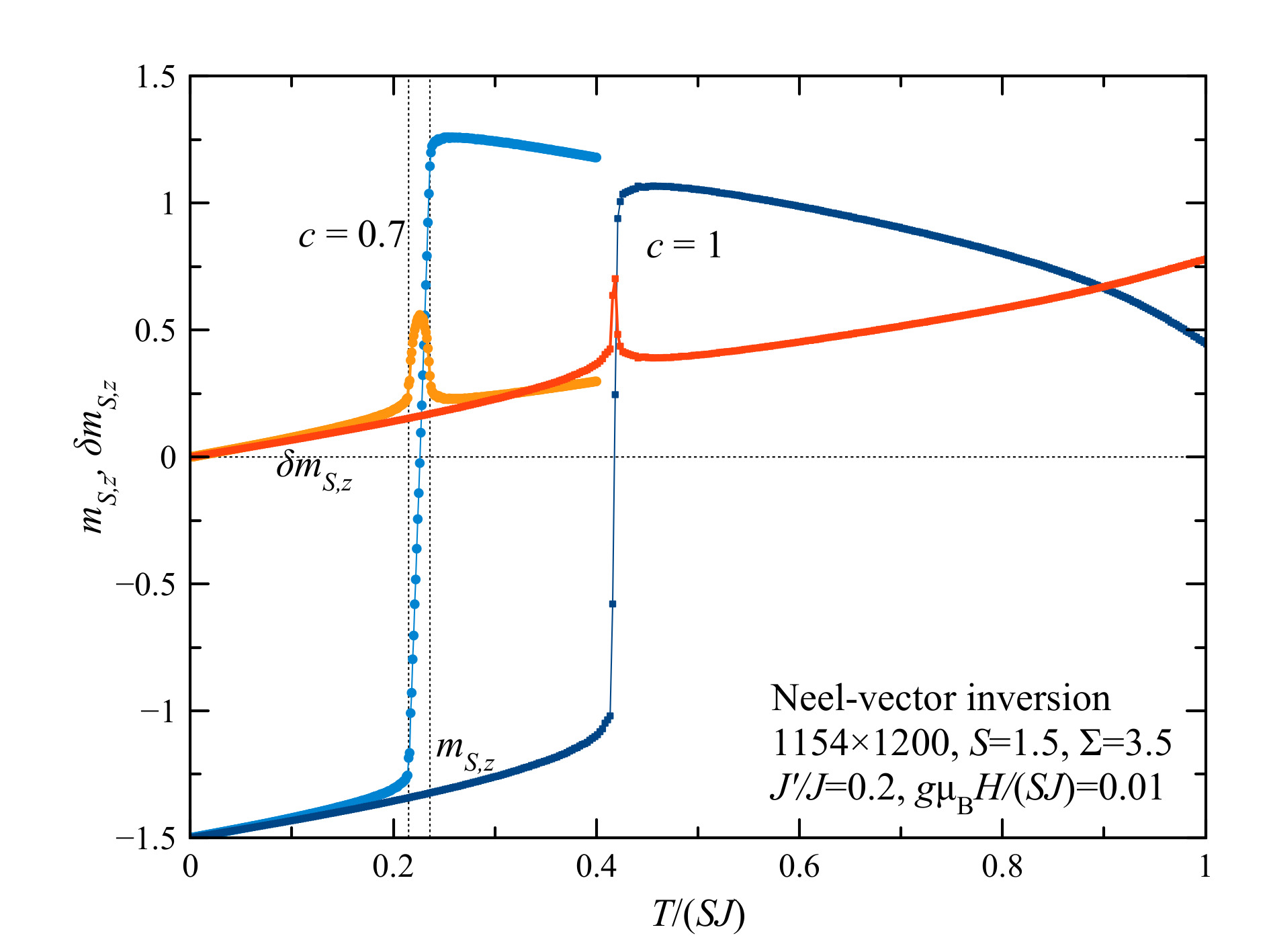}
\par\end{centering}
\begin{centering}
\includegraphics[width=8cm]{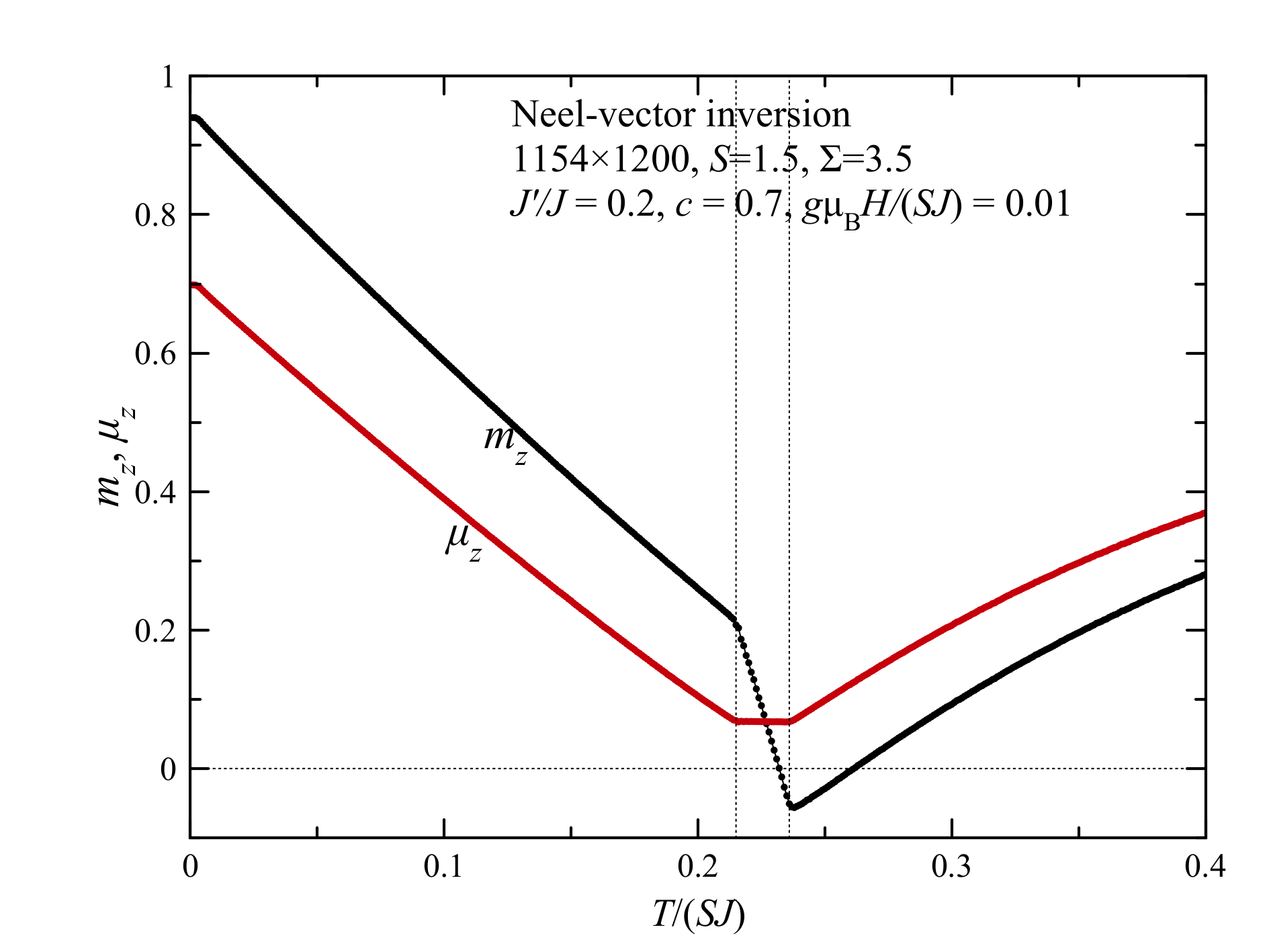}
\par\end{centering}
\caption{Spin-flip transition on temperature. Top: Angular momentum of the
TM sublattice, $m_{z}$, and dispersion of its fluctuations. Bottom:
Total angular momentum ($m_{z}$) and total magnetic moment ($\mu_{z}$).
Dotted vertical lines show the boundaries of the spin-flop phase for
$c=0.7$: $T/(SJ)=0.215$ and 0.236. }\label{Fig_Spin-flip-on_T}
\end{figure}

Figure \ref{Fig_Spin-flip-on_c}(bottom) shows $z$ component of the
average total spin defined by Eq. (\ref{m_vec_def}), as well as that
of the total magnetic moment normalized by that of the TM spins defined
by Eq. (\ref{mu_vec_def}), One can see that $\mu_{z}=\mathrm{const}>0$
in the transient region, while $m_{z}$ crosses the zero level twice.
Similar results were obtained for nonzero anisotropy, if $H>H_{+}$
given by Eq. (\ref{Hpm_at_compensation}). The intermediate spin-flop
phase was also studied in Ref. \citep{Haltz-PRB2022} within the mean-field
approximation.

The spin-flip transition on temperature is similar to that on the
RE concentration. The results for $c=0.7$ and $c=1$ in the case
of zero anisotropy are shown in Fig. \ref{Fig_Spin-flip-on_T}. Here
also, the intermediate spin-flop phase disordering, measured by $\delta m_{S,z}$,
is stronger than that of the surrounding collinear phase, even in
the absence of the local concentration fluctuations for the dense
ferrimagnet, $c=1.$ For $c=0.7$, the spin-flop phase exists in the
range $0.215\leq T/(SJ)\leq0.236$, between the dashed vertical lines
in the figure.

Let us rewrite these results in real units. Using $J=11\times10^{-22}$J
\citep{Berges2022}, $S=3/2$, and $g=2.2$, one obtains the existence
region of the spin-flop phase $25.7\mathrm{K}\leq T\leq28.2\mathrm{K}$
at the applied field $H=0.81$T. The MM compensation temperature,
according Eq. (\ref{mu_spin-flop_region}), is in the middle of this
narrow interval, that is, $T_{M}=26.0$K.

\section{Uniform excitation modes at different RE concentrations}

\label{Sec_Uniform-excitation-modes}

\begin{figure}
\begin{centering}
\includegraphics[width=8cm]{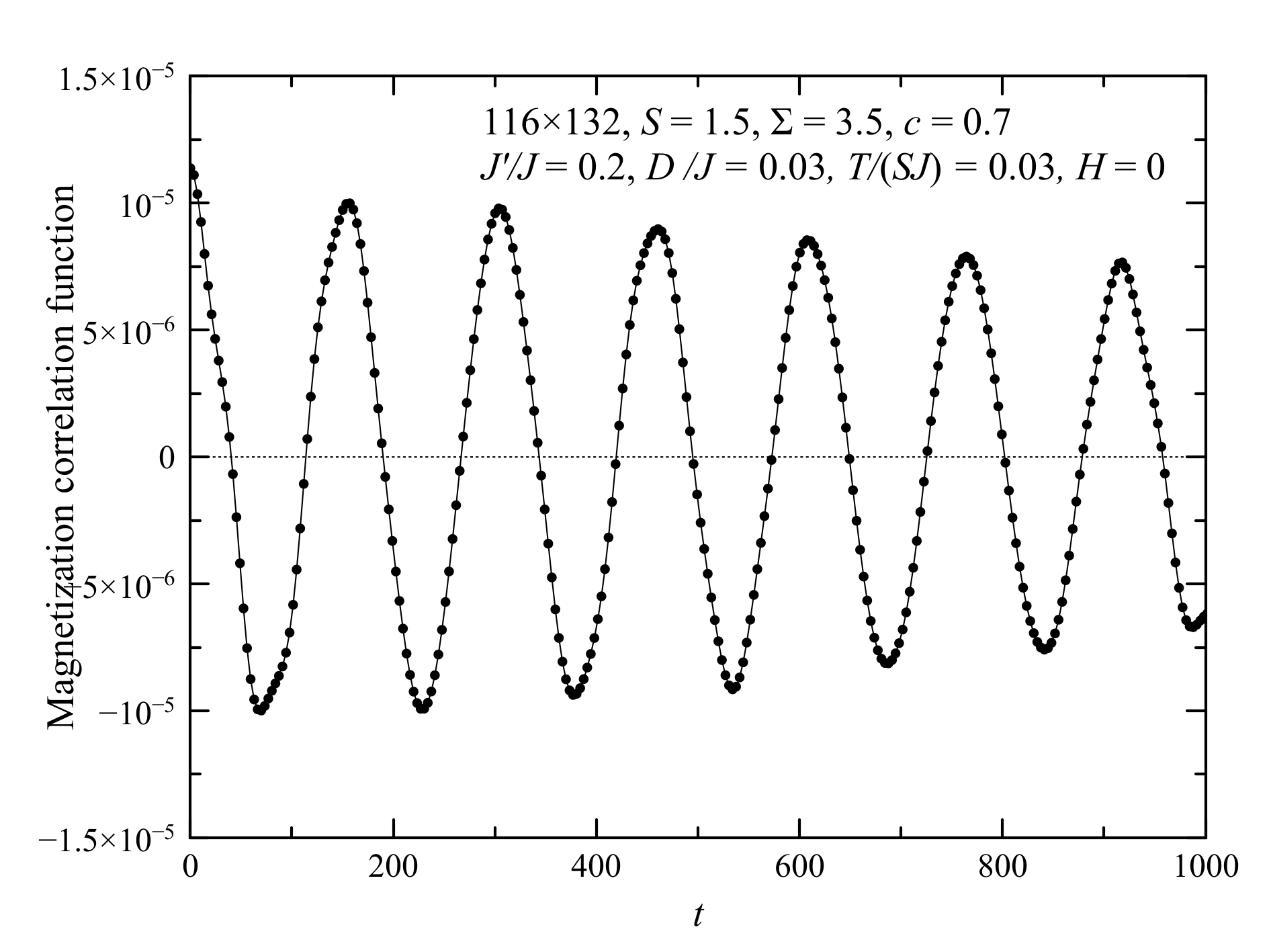}
\par\end{centering}
\begin{centering}
\includegraphics[width=8cm]{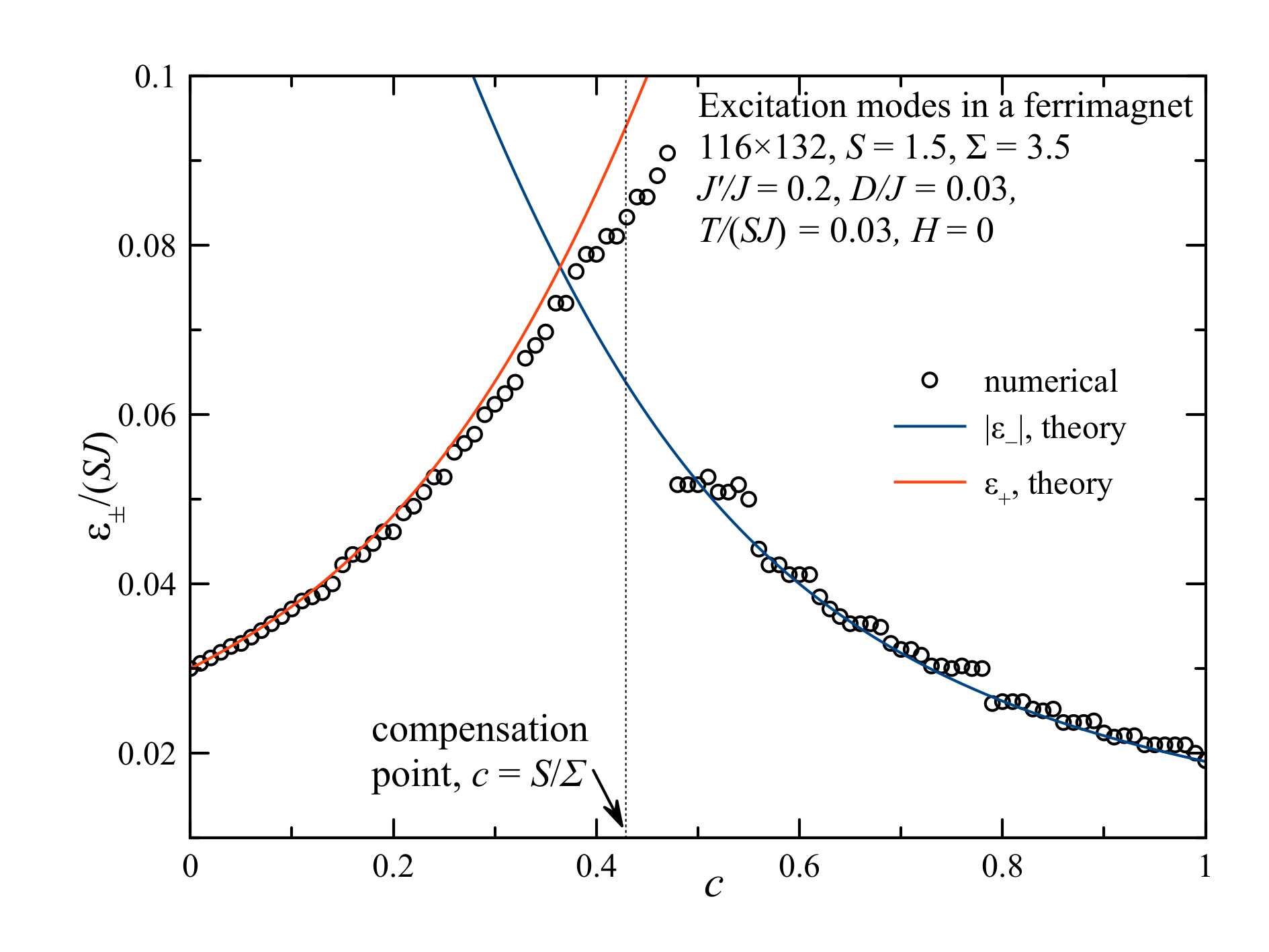}
\par\end{centering}
\caption{Energy spectrum of uniform modes at different RE concentrations, obtained
from the first period of the spin correlation function, compared with
the theory in Sec. \ref{Sec_Excitation-modes}.}\label{Fig_eps_from_one_period}
\end{figure}

\begin{figure}
\begin{centering}
\includegraphics[width=8cm]{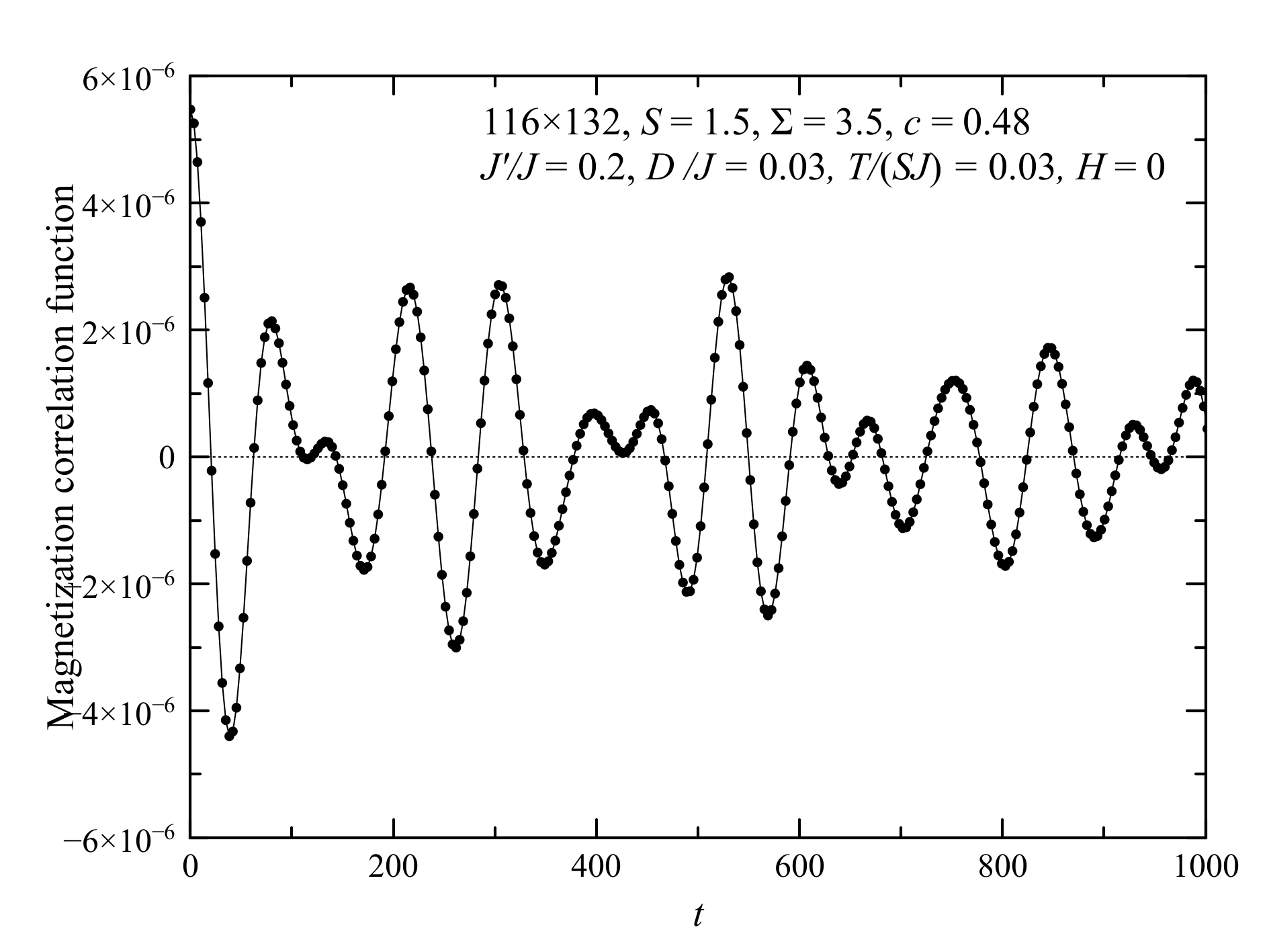}
\par\end{centering}
\caption{The magnetization CF close to the angular-momentum compensation point
(here $c=0.48)$ has a complicated form.}\label{Fig_CFt_c=00003D0.48}
\end{figure}

The dynamics of excitations in the uniform state of a ferrimagnet
is contained in the transverse magnetization time correlation function
(CF) defined by
\begin{equation}
A(t)=\frac{1}{2}\left(\left\langle \mu_{x}(t_{0})\mu_{x}(t_{0}+t)\right\rangle _{t_{0}}+\left\langle \mu_{y}(t_{0})\mu_{y}(t_{0}+t)\right\rangle _{t_{0}}\right).\label{CFt_def}
\end{equation}
To compute this CF, first the Landau-Lifshitz equation of motion is
solved numerically over a long interval of time starting from the
initial state at $t=0$ prepared by Monte Carlo equilibration at the
temperature $T$. Averaging over $t_{0}>0$ is computed as a self-correlation
of a list of data containing $N_{t}$ discrete time points, based
on the fast Fourier transform. (A direct straightforward computation
involves $N_{t}^{2}$ operations and is unfeasible). If one of the
two uniform modes dominates, one can extract the mode frequency from
the very first period of the CF. The results of this quick computation
are shown in Fig. \ref{Fig_eps_from_one_period}. The top graph shows
the magnetization CF at $c=0.7$ (far from the compensation point)
and $T/(SJ)=0.03$. In this case, the CF has a nice damped-sinusoidal
shape, and extracting the mode frequency from its first period is
unproblematic (the results are in the bottom panel of Fig. \ref{Fig_eps_from_one_period}).
A similar situation is everywhere not too close to the compensation
point. The frequency of the weaker mode cannot be found with this
method. Close to the compensation point, the two modes are comparably
strong, and the CF has a complicated shape as it is shown in Fig.
\ref{Fig_CFt_c=00003D0.48} for $c=0.48$. Also note that the amplitude
of this CF is smaller than for $c=0.7$, because the magnetic moment
of the system is small near the compensation point. At that point,
the method of extracting the frequency from the CF switches to another
mode, as can be seen in Fig. \ref{Fig_eps_from_one_period}(bottom).
Overall, this figure shows a good agreement with the analytical results
of Eq. (\ref{epsilon_pm-general}).

This simple method of obtaining the modes' frequencies and elucidating
the mode crossing at the magnetic compensation on temperature is problematic,
because the magnetic CF is distorted by thermal fluctuations at small
evolution times. Only at a sufficiently large evolution time the shape
of the CF stabilizes as fluctuations are averaged out. To obtain both
modes at any temperature, one needs to use a more fundamental approach
based of the fluctuation-dissipation theorem and compute the whole
absorption spectrum, see the next section.


\section{The absorption spectrum}

\label{Sec_The-absorption-spectrum}

\begin{figure}
\begin{centering}
\includegraphics[width=8cm]{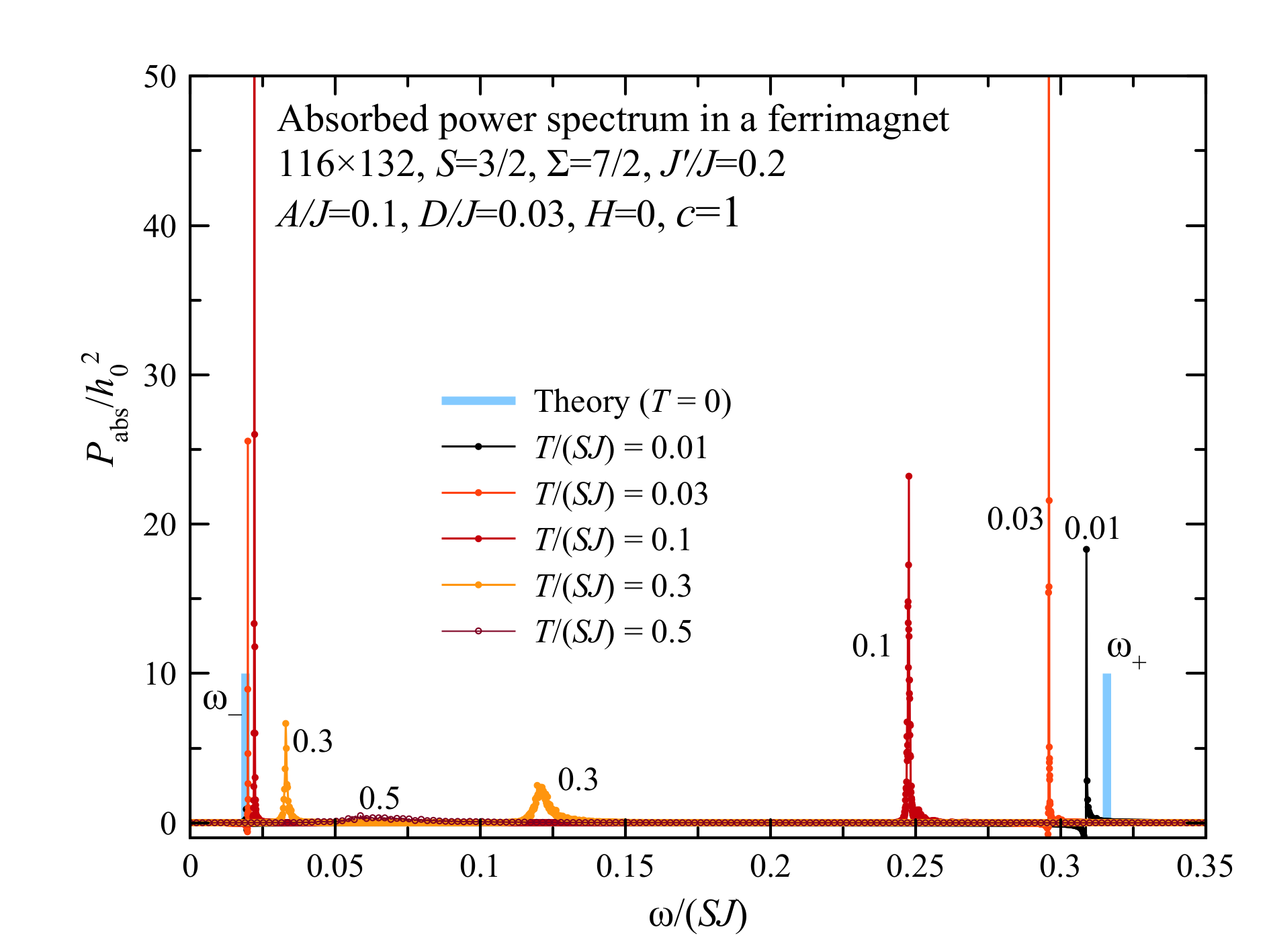}
\par\end{centering}
\caption{Absorption spectrum of an undiluted ferrimagnet with $J'/J=0.2$ at
different temperatures. With increasing $T$, the low-frequency mode
hardens and the high-frequency mode softens, until they meet at the
compensation point and broaden away {[}$T/(SJ)=0.5${]}.}\label{Fig_Pabs_undiluted_Jpr=00003D0.2}
\end{figure}
\begin{figure}
\begin{centering}
\includegraphics[width=8cm]{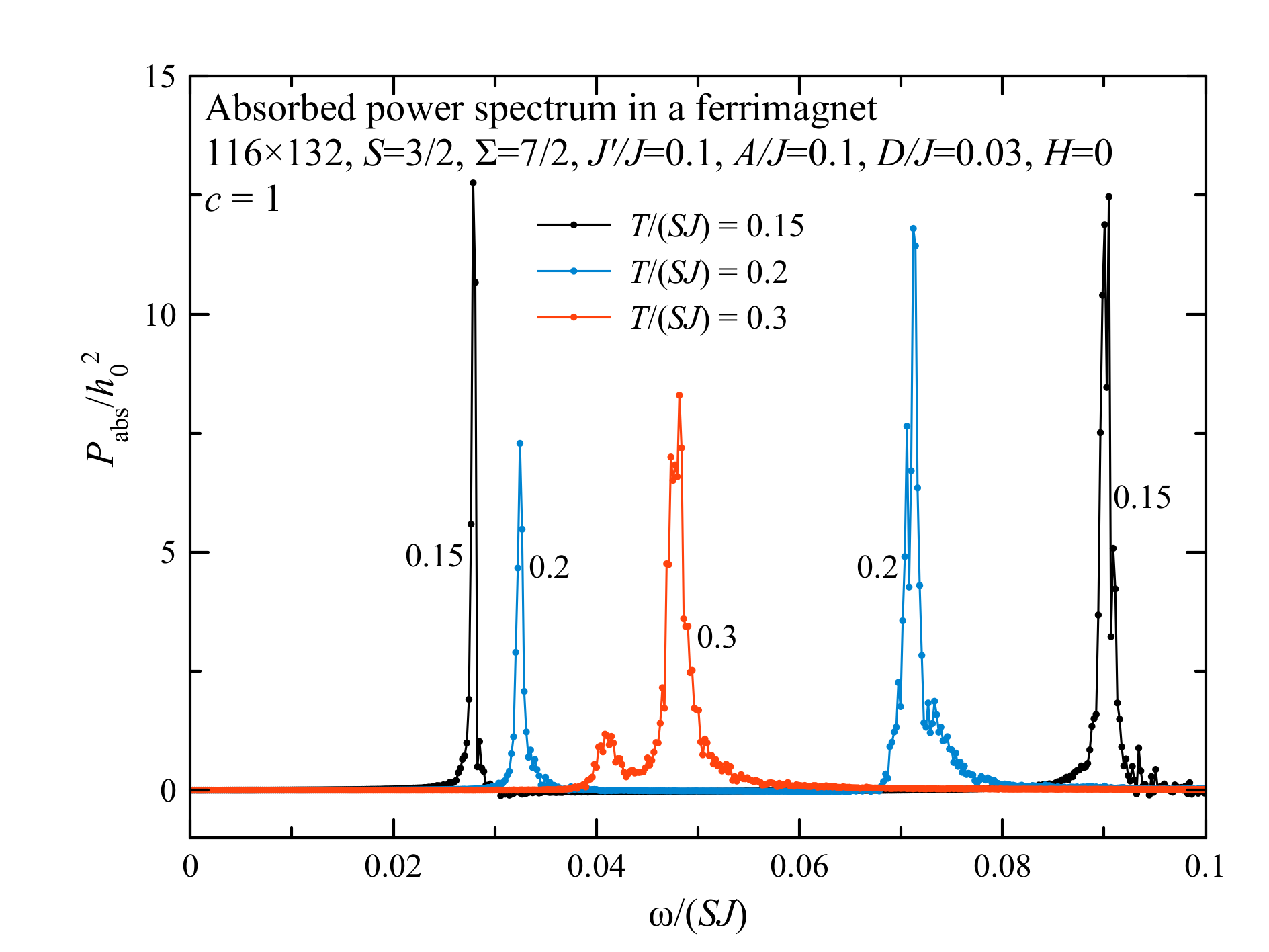}
\par\end{centering}
\begin{centering}
\includegraphics[width=8cm]{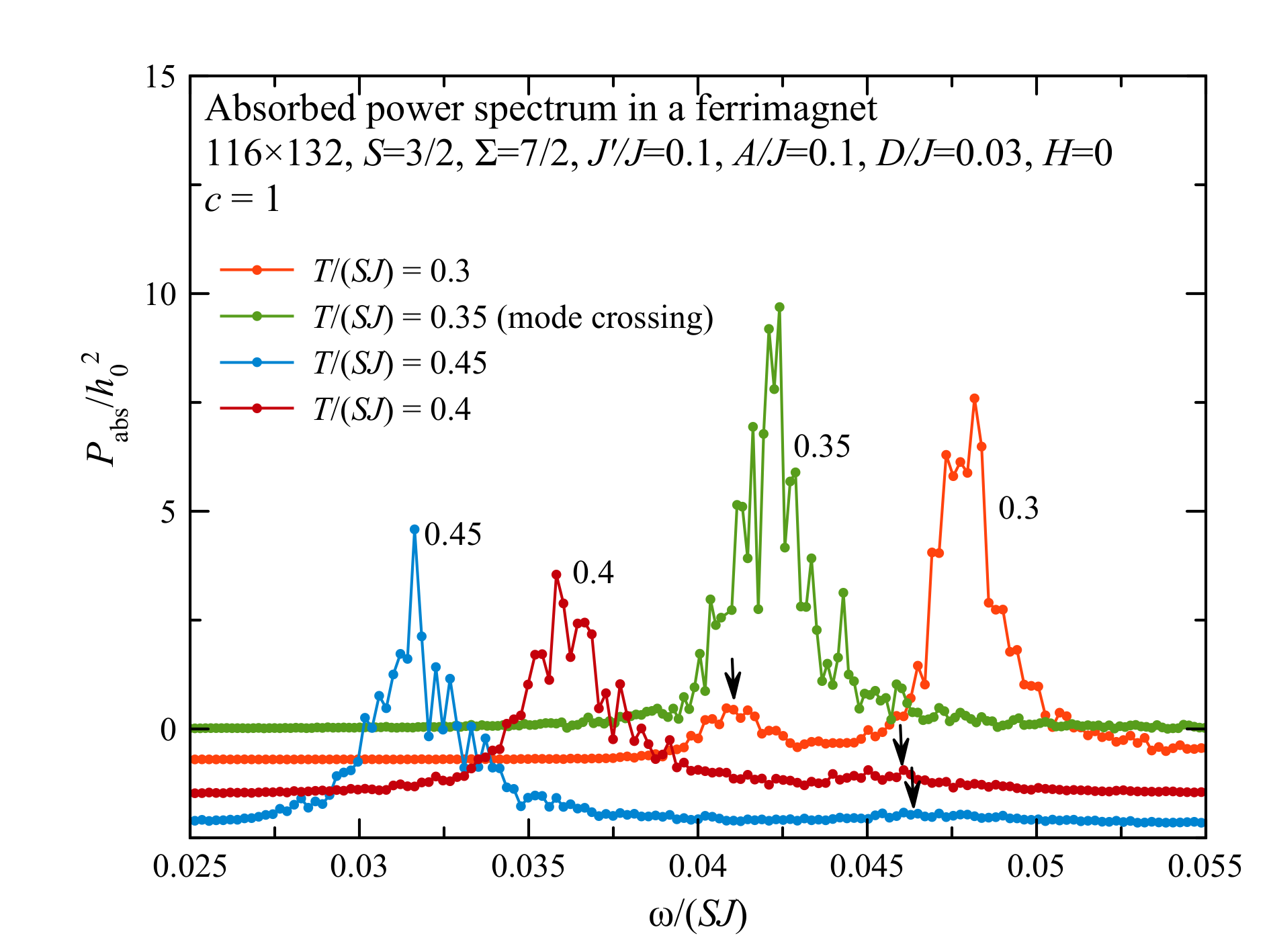}
\par\end{centering}
\caption{Absorption spectrum of an undiluted ferrimagnet with $J'/J=0.1$ at
different temperatures. Peak broadening is entirely due to intrinsic
processes in the magnetic system and increases with temperature. Top:
lower $T$ (two modes are well separated and clearly seen for each
$T$); Bottom: broad peaks at higher $T$ (vertical shifts added to
avoid overlapping and make satellites visible; for $T/(SJ)=0.3$ there
is a satellite on the left; for $T/(SJ)=0.4$, 0.45 there are very
small satellites on the right). }\label{Fig_Pabs_undiluted_Jpr=00003D0.1}
\end{figure}

To find the absorption spectrum, we compute the time dependence $\boldsymbol{\mu}(t)$,
Eq. (\ref{mu_vec_def}), over an extensive interval of time and define
the absorbed microwave (MW) power with the help of the fluctuation-dissipation
theorem (FDT) as
\begin{equation}
\frac{P_{\mathrm{abs}}(\omega)}{h_{0}^{2}}=\frac{\mathcal{N}\omega^{2}}{2k_{B}T}\mathrm{Re}\int_{0}^{\infty}dt\,e^{i\omega t}A(t),\label{P_abs_FDT}
\end{equation}
where $h_{0}$ is the amplitude of the MW radiation field in the energy
units and is the time correlation function of the transverse magnetization
components given by Eq. (\ref{CFt_def}). The frequency dependence
of $P_{\mathrm{abs}}(\omega)$ reflects the spectrum of uniform excitations
in the system. To find the frequencies of the modes, it is sufficient
to compute the dynamical evolution until the maximal time $t_{\max}$
that contains a large number of periods of the modes. A much longer
computation is needed to obtain accurate absorption peaks with well-defined
widths. Absorption peaks can be considered as well-formed when $P_{\mathrm{abs}}(\omega)>0$
everywhere. This can be checked as the computation runs to define
the required end time $t_{\max}$.

The absorption spectrum of an undiluted ferrimagnet with the intersublattice
coupling constant $J'/J=0.2$ at different temperatures is shown in
Fig. \ref{Fig_Pabs_undiluted_Jpr=00003D0.2}. There are two peaks
at each temperature, corresponding to the two uniform modes. There
is a good agreement with the theoretical values of the frequencies
given by Eq. (\ref{epsilon_pm-general}), which was used to calculate
the modes' energies at $T=0$. The softening of the high-frequency
mode with temperature is due to the thermal disordering of the weakly
coupled RE spins. Note that the HF peak is not well-formed at $T/(SJ)=0.01$,
and there is a region of a negative $P_{\mathrm{abs}}$. This is because
the damping of this mode is very low and the dynamical evolution time
was insufficient for the time correlation function to approach zero.
The same applies to the LF modes at low temperatures, which have a
lower damping, than the HF mode. Still, even non-well-formed peaks
correctly show the resonance frequencies. With increasing $T$, absorption
peaks broaden. The likely mechanism for this is thermal disordering
of weakly-coupled RE spins. The latter act as random fields applied
to the more thermally stable TM subsystem. Note that we do not add
any phenomenological damping to the theory.

As for $J'/J=0.2$ in our model compensation occurs at a rather high
temperature, the two modes are already strongly damped at mode crossing
{[}$T/(SJ)=0.5${]}, and further temperature increase makes them broadened
out and unobservable. For the lower intersublattice coupling, $J'/J=0.1$,
compensation occurs at lower temperatures, so that the mode-crossing
can be observed, in spite of the peaks being broad, too. The evolution
of the absorption spectrum with increasing the temperature is shown
in Fig. \ref{Fig_Pabs_undiluted_Jpr=00003D0.1}. Above the compensation,
there is a broad LF peak and a very small broad HF satellite, whose
frequency (energy) can still be determined.

\begin{figure}
\begin{centering}
\includegraphics[width=8cm]{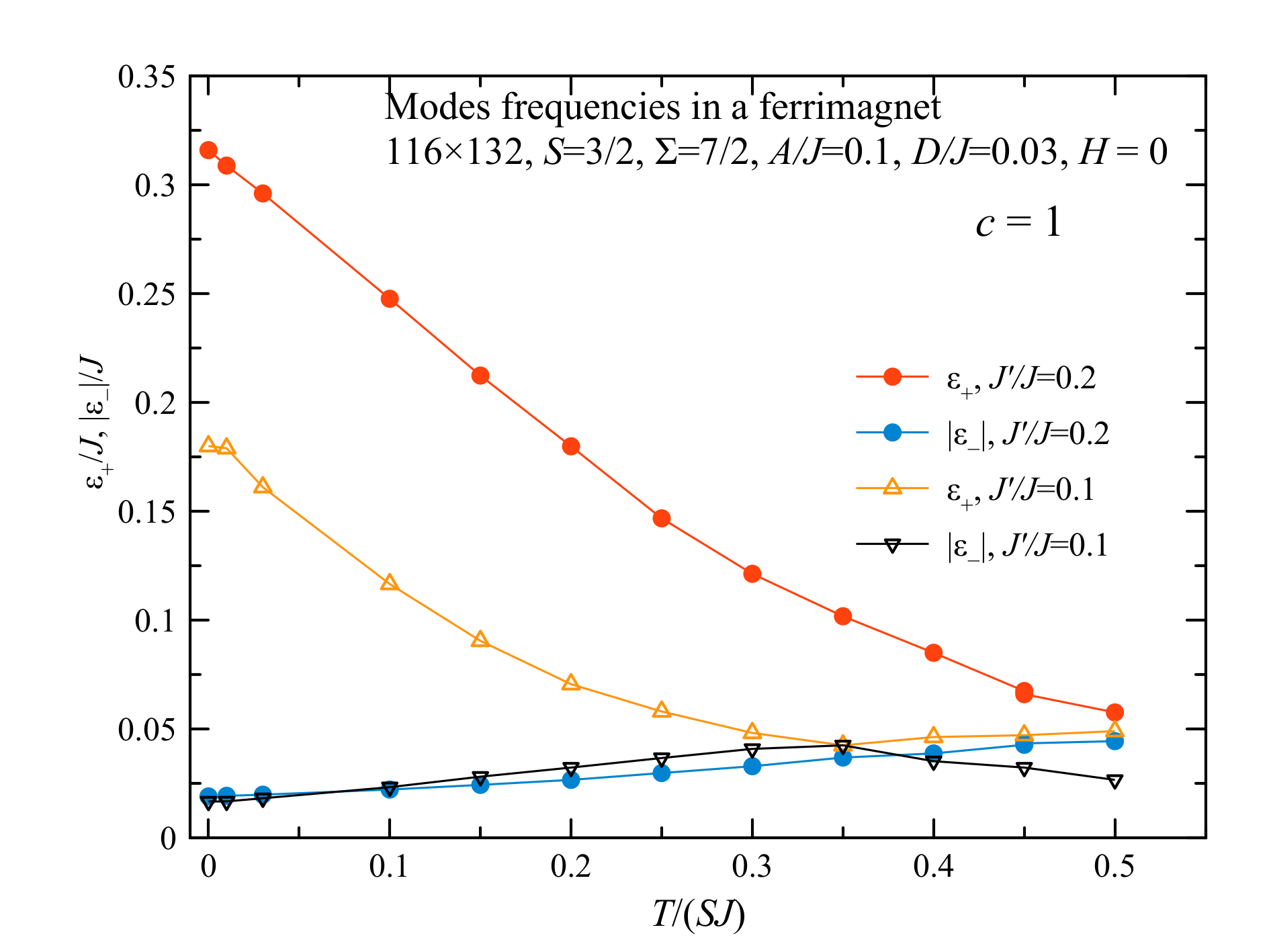}
\par\end{centering}
\caption{Temperature dependence of the uniform ferrimagnetic modes of an undiluted
ferrimagnet with $J'/J=0.2$ and 0.1. The values at $T=0$ are calculated
analytically using Eq. (\ref{epsilon_pm-general}). For $T/(SJ)\protect\geq0.5$,
the peaks in the absorption spectrum broaden out.}\label{Fig_Temperature-dependence-of_modes}
\end{figure}

Temperature dependences of the energies of the uniform modes of the
undiluted system extracted from the absorption spectra are shown in
Fig. \ref{Fig_Temperature-dependence-of_modes} both for $J'/J=0.2$
and $J'/J=0.1$. As said above, the mode crossing is visible only
for $J'/J=0.1$.

\begin{figure}
\begin{centering}
\includegraphics[width=8cm]{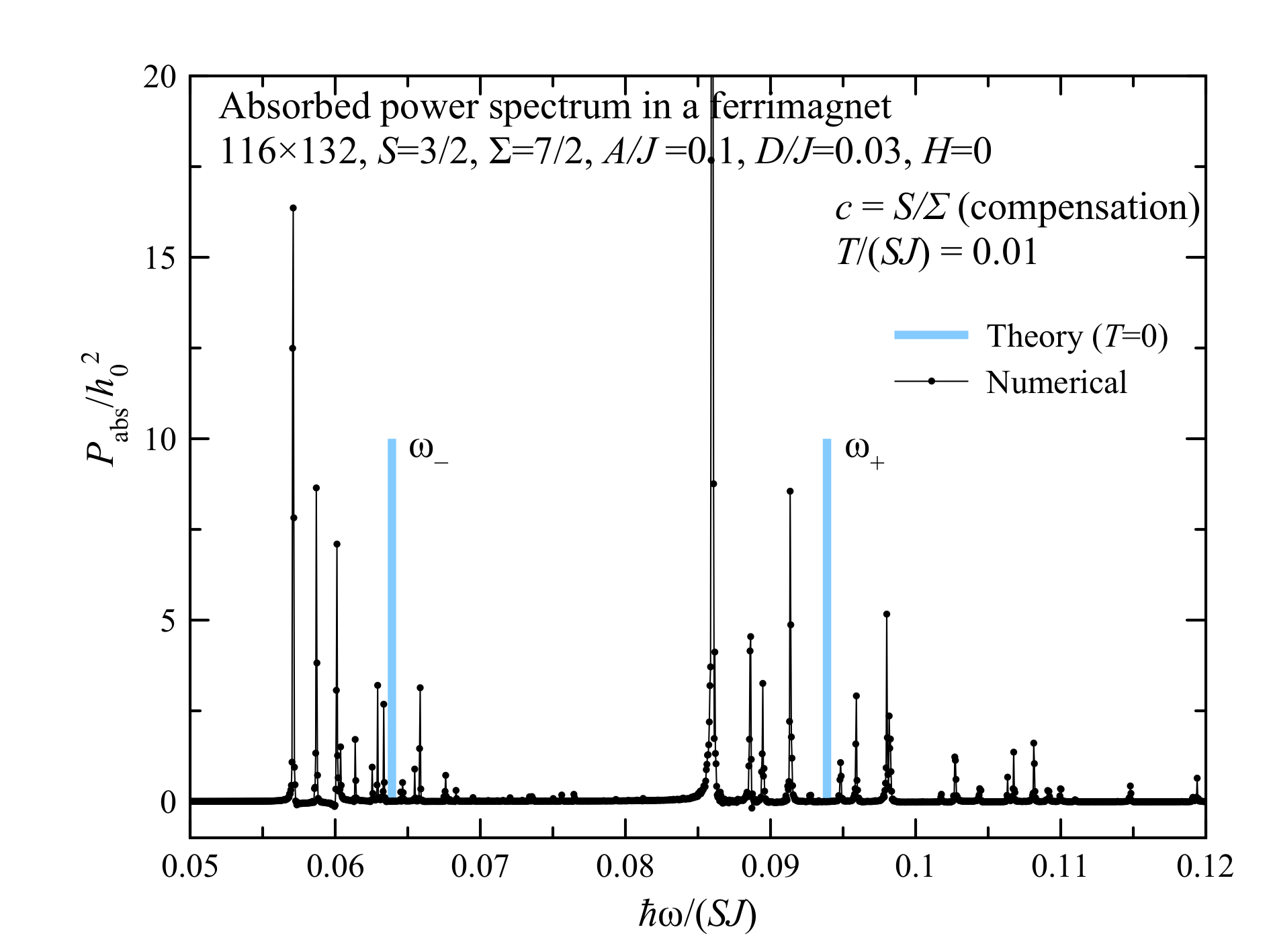}
\par\end{centering}
\begin{centering}
\includegraphics[width=8cm]{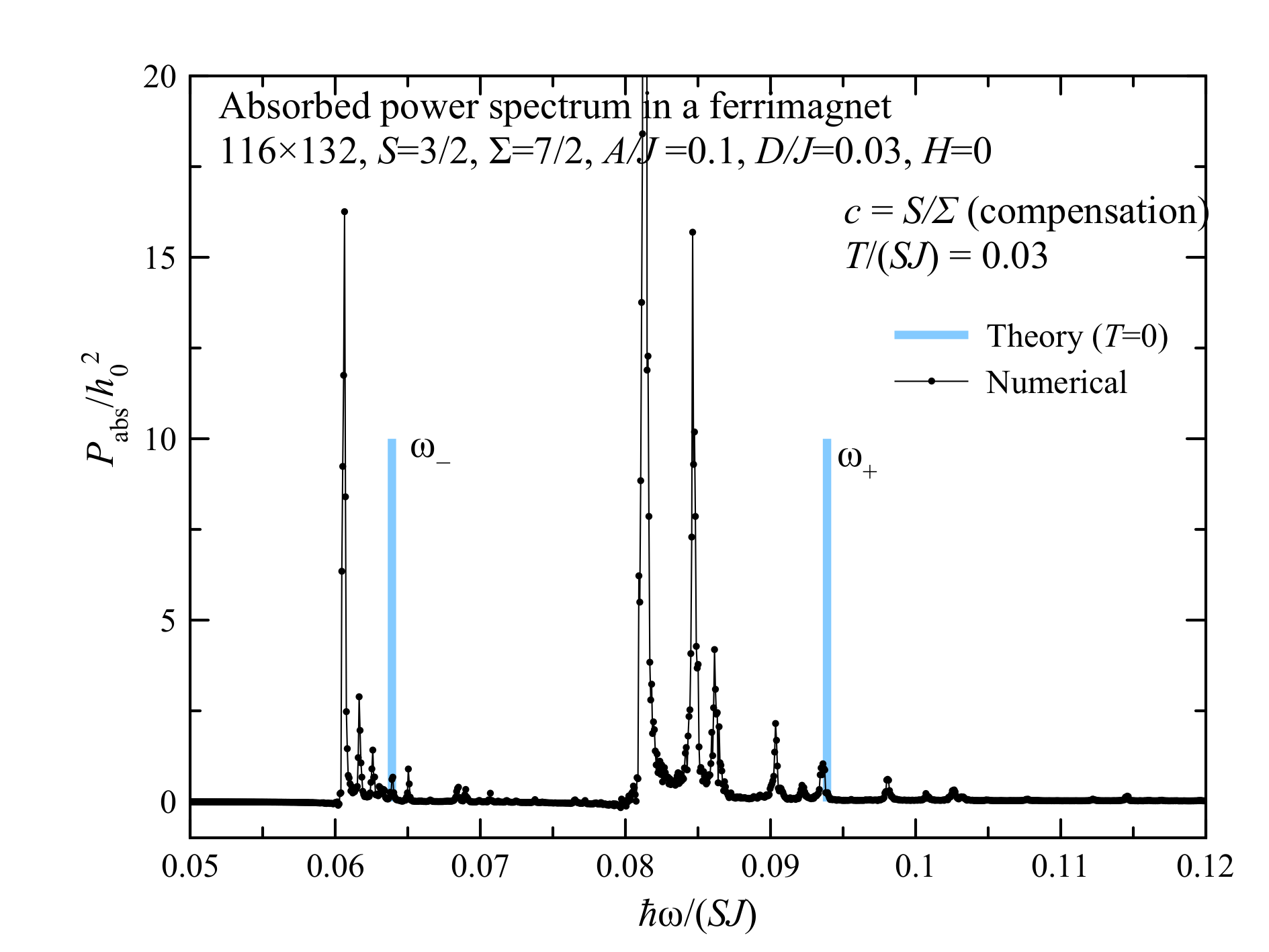}
\par\end{centering}
\caption{The absorption spectrum of a diluted ferrimagnet at the angular-momentum
compensation point at low temperatures, $T/(SJ)=0.01$ and 0.03. The
peaks in these plots, corresponding to local modes, do not coincide
because of different realizations of the spatial distribution of RE
atoms. }\label{Fig_Pabs_compensation}
\end{figure}

The dilution of RE atoms adds a new feature to the absorption spectrum,
in addition to the temperature dependences of the uniform modes. Fig.
\ref{Fig_Pabs_compensation} shows peaks split into several sub-peaks.
This is a consequence of static disorder in the diluted ferrimagnet.
The RE atoms are not uniformly distributed, and their local concentration
fluctuates. As a result, the translational invariance is broken, and
instead of a single uniform mode, there are many localized modes with
different frequencies in the system. A similar behavior in ferromagnets
with random anisotropy was seen in Ref. \citep{GarChu2023}. The positions
of the peaks in the plots for $T/(SJ)=0.01$ and 0.03 are different
because of different realizations of the spatial distribution of RE
atoms.

\begin{figure}
\begin{centering}
\includegraphics[width=8cm]{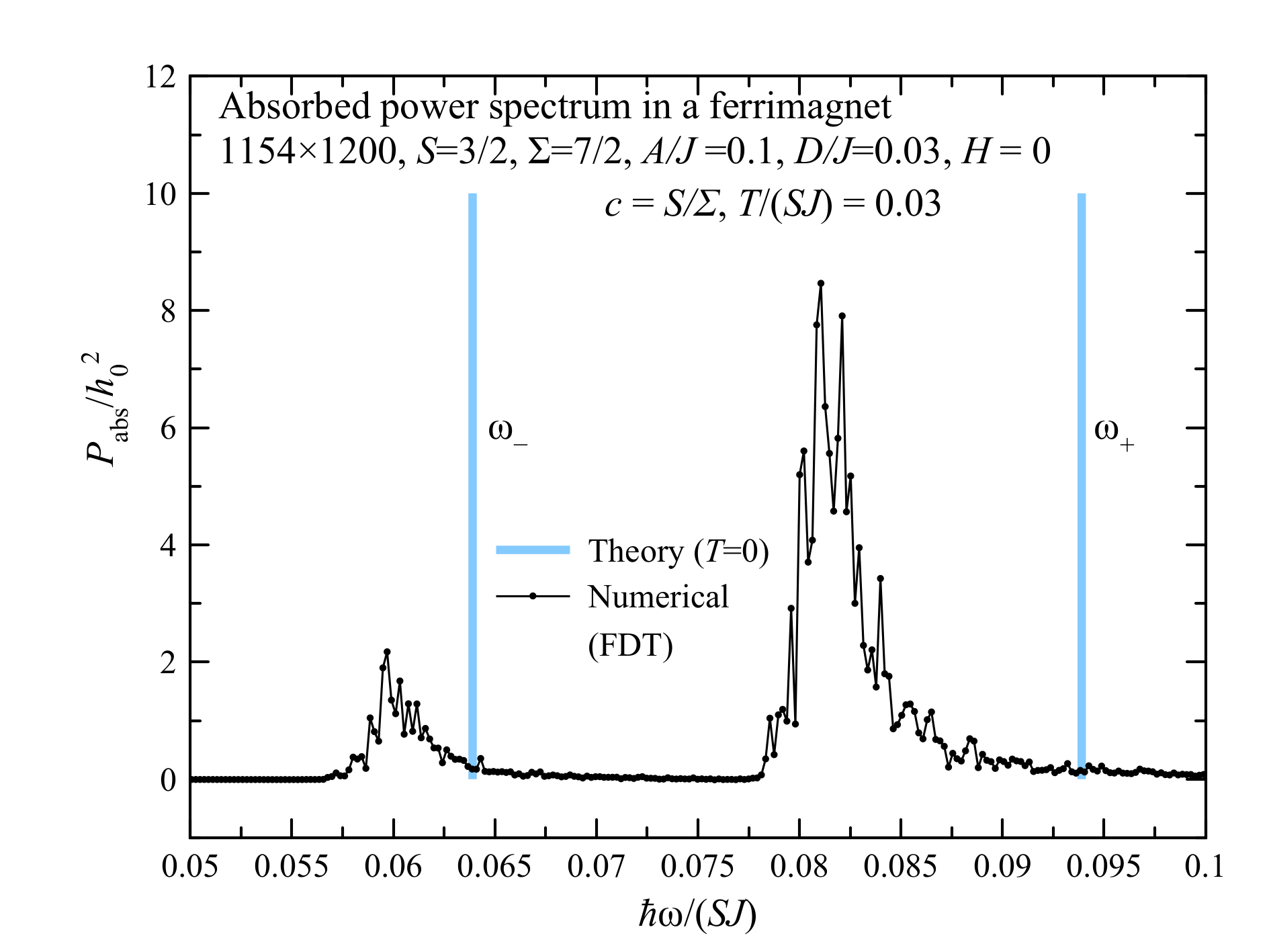}
\par\end{centering}
\caption{The absorption spectrum of a diluted ferrimagnet at the angular-momentum
compensation point at low temperatures, $T/(SJ)=0.03$, in a larger
system of $1154\times1200$ lattice sites. Narrow peaks corresponding
to the local modes, seen in the preceding graphs, here merge into
broad peaks. }\label{Fig_Pabs_compensation_larger}
\end{figure}

The system of $116\times132$ lattice sites in the figures above is
still too small, so that different localized modes are split. In large
systems, localized modes build a continuum, that is, broad damped
peaks even at $T=0$. This can be seen in Fig. \ref{Fig_Pabs_compensation_larger}
for the system of $1154\times1200$ lattice sites, about 100 times
larger than in the $116\times132$ system. This computation took about
a month of computer time, with $t_{\mathrm{max}}\approx10^{6}$ in
computing units. One interesting feature of the broad peaks is their
asymmetry, with a tail extending towards high frequencies, which is
a general feature of systems with static randomness (see, e.g., Figs.
3-6 of Ref. \citep{garchu21prb}) that can be tested in experiments.

It is difficult to extract the temperature and concentration dependences
of the modes' frequencies of a diluted system from the absorption
spectrum. The problem is that for small systems, such as $116\times132$
lattice sites, one can perform the dynamical evolution and obtain
the spectrum for many values of $T$ or $c$, but the spectrum consists
of many peaks due to the modes' localization, as shown in Fig. \ref{Fig_Pabs_compensation},
and extraction of the peak frequency is problematic. To obtain a single
peak for each of the two modes, one needs to use much larger systems,
as in Fig. \ref{Fig_Pabs_compensation_larger}, This makes the extraction
of the peaks' frequencies straightforward, but such computations for
many values of $T$ or $c$ would take too much time.

\section{The integral absorption}

\label{Sec_The-integral-absorption}

\begin{figure}
\begin{centering}
\includegraphics[width=8cm]{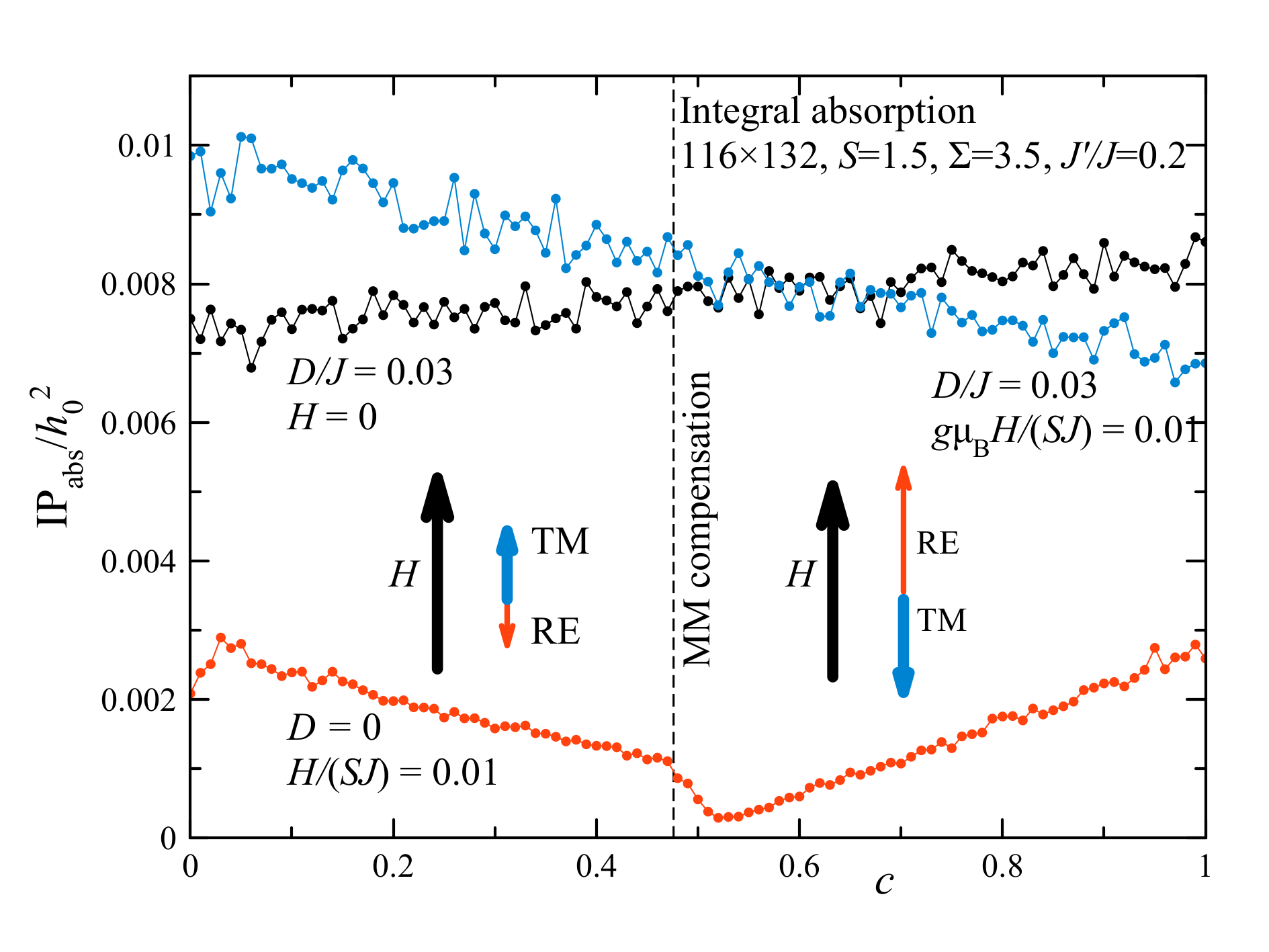}
\par\end{centering}
\begin{centering}
\includegraphics[width=8cm]{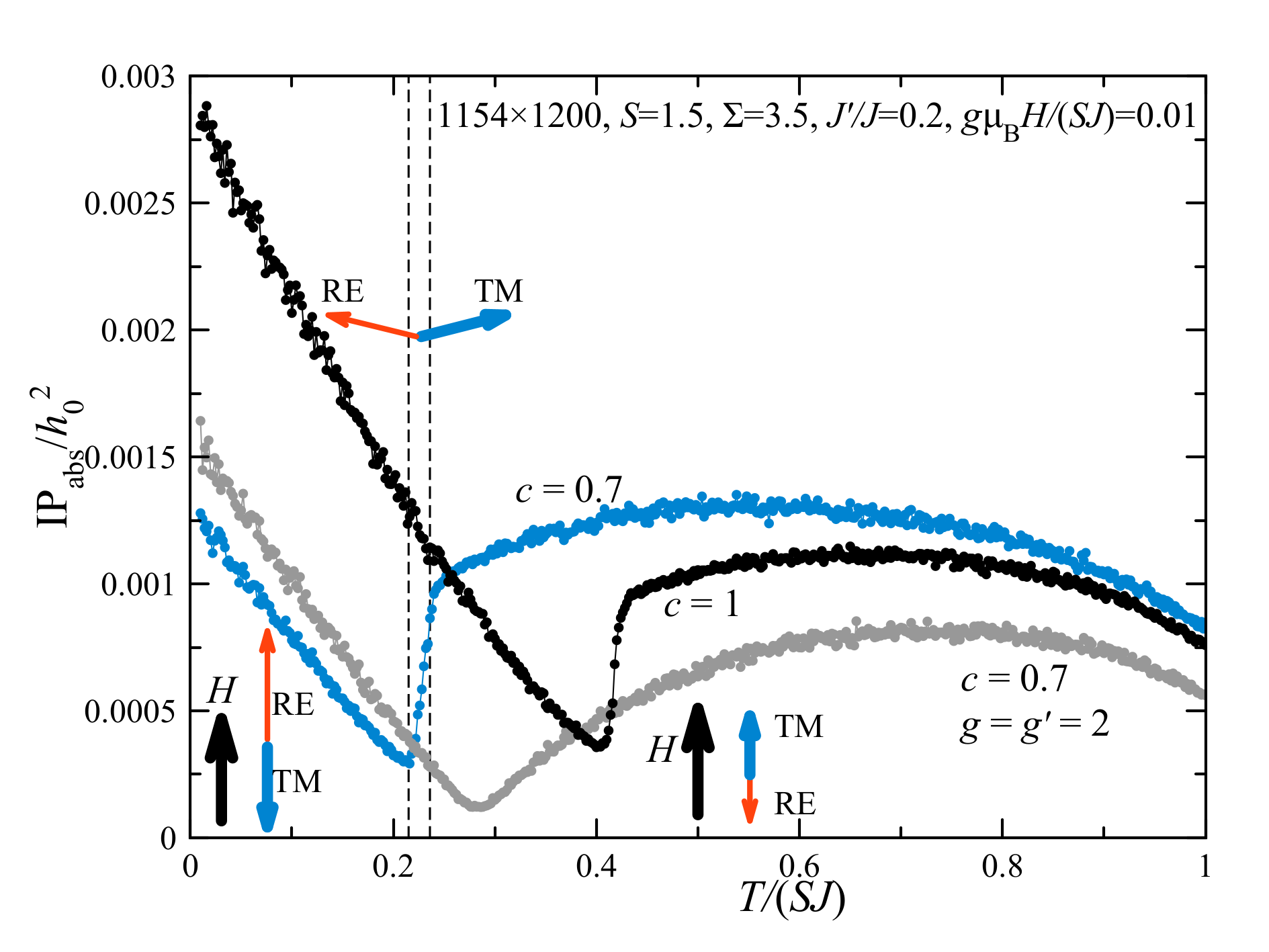}
\par\end{centering}
\caption{The RE concentration and temperature dependences of the integral absorption
in a ferrimagnet. Top: the RE concentration dependence with and without
anisotropy at zero and non-zero fields. For $D=0$, there is a minimum
at the spin-flip transition. Bottom: the temperature dependence across
the spin-flop transition for $D=0$. The blue and black curves are
for our main model with $g=2.2$ and $g'=2$, while the gray curve
is for the model with the same gyromagnetic ratio for the two sublattices,
$g=g'=2.$}\label{Fig_IPabs}
\end{figure}

The integral absorbed power defined as
\begin{equation}
IP_{\mathrm{abs}}=\int_{-\infty}^{\infty}\frac{d\omega}{2\pi}P_{\mathrm{abs}}(\omega)\label{IPabs_def}
\end{equation}
is a useful measure of the ability of a material to absorb the microwave
power \citep{ChuGar2023}. It depends on the strengths of the magnetic
resonances in the body, and, e.g., for material with the coherent
anisotropy $D$ one has $IP_{\mathrm{abs}}\propto Dh_{0}^{2}$, where
$h_{0}$ is the amplitude of the microwave field. By contrast, for
the model with random anisotropy of the strength $D_{R}$, one obtains
$IP_{\mathrm{abs}}\propto\left(D_{R}^{2}/J\right)h_{0}^{2}$, both
in 2D and 3D. The formula for the absorbed power, averaged over all
directions, has the form \citep{ChuGar2023}
\begin{equation}
\frac{IP_{\mathrm{abs}}}{h_{0}^{2}}=\frac{\mathcal{N}}{12k_{B}T}\left\langle \mathbf{\dot{M}}(t)^{2}\right\rangle _{t},\label{IPabs_via_M}
\end{equation}
where $\mathbf{\dot{M}}$ is the time derivative of the magnetic moment
of the system per lattice site and $h_{0}$ is in magnetic units (Tesla).
This can be rewritten in terms of the normalized magnetic moment $\boldsymbol{\mu}$
defined by Eq. (\ref{mu_vec_def}) as
\begin{equation}
\frac{IP_{\mathrm{abs}}}{h_{0}^{2}}=\frac{\mathcal{N}}{12k_{B}T}\left\langle \boldsymbol{\dot{\mu}}(t)^{2}\right\rangle _{t},\label{IPabs_via_mu_vec}
\end{equation}
where now $h_{0}$ is in the energy units. Using the equations of
motion for the TM and RE spins, Eq. (\ref{Larmor}), one further transforms
this formula to
\begin{equation}
\frac{IP_{\mathrm{abs}}}{h_{0}^{2}}=\frac{\left\langle \left[\sum_{i}\left(\mathbf{S}_{i}\times\mathbf{H}_{\mathrm{eff},i}+\frac{g'}{g}\boldsymbol{\sigma}_{i}\times\mathbf{H}'_{\mathrm{eff},i}\right)\right]^{2}\right\rangle _{t}}{12\mathcal{N}\hbar^{2}k_{B}T}.\label{IPabs_via_Larmor}
\end{equation}
Originally, averaging in this formula is performed over the dynamical
evolution of the system. However, it can be replaced by the averaging
over the Monte Carlo process, which greatly simplifies the computation.
It turns out that there is no self-averaging in large systems, so
one needs to perform a long Monte-Carlo evolution to average out the
noise.

It is interesting to see what is the influence of the added RE spins
on the MW absorption. The first thought is that more spins will give
more absorption. The second thought is that there should be less absorption,
because the TM and RE spins are coupled antiferromagnetically and
the total magnetization is reduced, especially near the compensation
point.

The results for the integral absorption are shown in Fig. \ref{Fig_IPabs}.
In the case of a sufficiently strong anisotropy preventing a spin
flip transition (the cases of $D/J=0.03$ and $H/(SJ)=0,\,0.01$)
there is no influence of the compensation on the integral absorption,
as can be seen from the concentration dependences in Fig. \ref{Fig_IPabs}(top).
For $H=0$, the integral absorption is nearly a horizontal line. For
$D=0$, the system undergoes the spin-flip transition at the magnetic-moment
compensation point (see Figs. \ref{Fig_Spin-flip-on_c} and \ref{Fig_Spin-flip-on_T}),
and here the integral absorption has a minimum. A similar minimum
is seen in the temperature dependence in the bottom panel of Fig.
\ref{Fig_IPabs}. Note that flipping the TM sublattice into the direction
of the applied field greatly increases the absorption. The dip in
$IP_{\mathrm{abs}}$ is so asymmetric because in our model, $g>g$'.
This allows to have a significant magnetization {[}see Fig. \ref{Fig_Spin-flip-on_T}(bottom){]}
and significant absorption close to the angular-momentum compensation
point, on one of the sides of the spin-flip transition. In the model
with $g=g',$ there is a symmetric minimum. A take-away of these investigations
is that not the compensation per se but the spin-flip transition creates
a nontrivial behavior of the integral absorption.

\section{Conclusions}

We have developed numerical methods for an efficient computation of
temperature and composition dependence of equilibrium magnetic states,
uniform oscillation modes, and microwave power absorption in transition-metal/rare-earth
ferrimagnetic alloys. They include energy minimization on large spin
lattices, adaptive Monte Carlo routine with thermalized overrelaxation,
and the numerical solution of the dynamics of conservative spin systems
at nonzero temperatures robust at any large times and specific to
ferrimagnets.

Our low temperature results on the temperature dependence of magnetic
resonance modes and their hardening on approaching the angular momentum
compensation point agree well with the analytical theory for ferrimagnets
developed in the past \citep{Wangsness-PR1953,Lin-PRB1988,Zhang-JPhys1997,Karchev-JPhys2008,Okuno-APLExpress2019,Haltz-PRB2022,Sanchez-PRB2025}
and presented for our model in Section II. The dependence of the magnetization
on the temperature and concentration of rare-earth atoms has been
computed. In agreement with previous theoretical work and experiments
(see, e.g., Refs. {[}\onlinecite{Binder-PRB2006,Stanciu-PRB2006,Joo-materials2021,Chanda-PRB2021,Haltz-PRB2022}{]}),
it exhibits re-orientation of the magnetizations of sublattices close
to angular momentum and magnetic moment compensation points. Large
fluctuations of the magnetization have been observed in the vicinity
of the compensation.

By computing the magnetization correlation function and utilizing
the fluctuation-dissipation theorem, we obtained the frequency dependence
of the absorbed microwave power at different temperatures and compositions
of the transition metal/rare-earth ferrimagnet. Static disorder due
to random positions of rare-earth atoms in the transition metal matrix
results in peculiar features in the spectrum of spin oscillations.
Multiple resonances have been observed in the power absorption that
correspond to spatially localized excitations. As the size of the
system increases, they produce a broad excitation band close to the
analytically computed resonances for an ideal system of two antiferromagnetically
coupled ferromagnetic sublattices. The integral of the absorbed power
over frequency exhibits a minimum near the angular momentum compensation
point.

Our analytical results apply to both 2D and 3D. The numerical results
are obtained in 2D, keeping in mind thin films. Since this is not
a first-principles calculation, we do not pretend to provide exact
numbers for comparison with experiments, aiming at a qualitative description
instead. For the class of systems that we are studying, the difference
in the coordination numbers for the TM and RE atoms in 2D and 3D can
only alter the results by factors of order unity. A detailed study
involving the coordination numbers for a ferrimagnetic system like,
e.g., the $\mathrm{Co}_{1-x}\mathrm{Gd}_{x}$ intermetallic compound,
would be more involved because of the dependence of the stoichiometry
on the composition, with only several unique compositions having a
well-defined crystal structure. Introducing random positions of the
RE atoms in ~a lattice is the most reasonable approximation one can
make when studying magnetic properties of the TM/RE compound over
a wide range of RE concentration.

While the method described in this paper was developed for a two-sublattice
ferrimagnet and used parameters of CoGd ferrimagnetic alloys, it can
be applied to any ferrimagnetic system, with most of our conclusions
expected to remain qualitatively valid.

\section*{Acknowledgements}

This work has been supported by Grants No. FA9550-24-1-0090 and FA9550-24-1-0290
funded by the Air Force Office of Scientific Research.

\section*{Appendix: ferrimagnet's excitation modes}

Here we present a derivation of the excitation modes in a ferrimagnet.
Consider small deviations from the anticollinear state aligned along
the $z$ axis, $S_{iz}=S$ and $\sigma_{iz}=-\Sigma$ in the model
with $\mathbf{H}=H\mathbf{e}_{z}$ and linearize the equations of
motion in small transverse components of the spins:
\begin{eqnarray}
\mathbf{S}_{i} & = & \mathbf{S}_{i}^{(0)}+\mathbf{S}_{i}^{(1)},\nonumber \\
\mathbf{S}_{i}^{(0)} & = & S\mathbf{e}_{z},\qquad\mathbf{S}_{i}^{(1)}=S_{ix}\mathbf{e}_{x}+S_{iy}\mathbf{e}_{y},
\end{eqnarray}
etc. For the TM spins one obtains
\begin{equation}
\mathbf{\dot{S}}_{i}^{(1)}=\frac{1}{\hbar}\left[S\mathbf{e}_{z}\times\mathbf{H}_{\mathrm{eff},i}^{(1)}\right]+\frac{1}{\hbar}\left[\mathbf{S}_{i}^{(1)}\times\mathbf{H}_{\mathrm{eff},i}^{(0)}\right],
\end{equation}
where
\begin{eqnarray}
\mathbf{H}_{\mathrm{eff},i}^{(0)} & = & \left[S\left(J_{0}+D\right)+g\mu_{B}H+p_{i}\Sigma J'\right]\mathbf{e}_{z}\nonumber \\
\mathbf{H}_{\mathrm{eff},i}^{(1)} & = & \sum_{j}J_{ij}\mathbf{S}_{j}^{(1)}-J'p_{i}\boldsymbol{\sigma}_{i}^{(1)}
\end{eqnarray}
and $J_{0}=zJ$, $z$ is the number of the nearest neighbors.

For the RE spins one obtains
\begin{equation}
\dot{\boldsymbol{\sigma}}_{i}^{(1)}=\frac{1}{\hbar}\left[-\Sigma\mathbf{e}_{z}\times\mathbf{H}'{}_{\mathrm{eff},i}^{(1)}\right]+\frac{1}{\hbar}\left[\boldsymbol{\sigma}_{i}^{(1)}\times\mathbf{H}'{}_{\mathrm{eff},i}^{(0)}\right],
\end{equation}
where
\begin{eqnarray}
\mathbf{H}'{}_{\mathrm{eff},i}^{(0)} & = & \left(-\Sigma D_{\Sigma}+g'\mu_{B}H-J'S\right)\mathbf{e}_{z}\nonumber \\
\mathbf{H}'{}_{\mathrm{eff},i}^{(1)} & = & -J'\mathbf{S}_{i}^{(1)}.
\end{eqnarray}

As soon as the equations of motion are linearized, one can perform
the Fourier transformation:
\begin{equation}
\mathbf{S}_{\mathbf{k}}=\frac{1}{\mathcal{N}}\sum_{i}\mathbf{S}_{i}e^{i\mathbf{k}\cdot\mathbf{r}_{i}},\qquad\mathbf{S}_{i}=\sum_{\mathbf{k}}\mathbf{S}_{\mathbf{k}}e^{-i\mathbf{k}\cdot\mathbf{r}_{i}}.
\end{equation}
For the parent sublattice one has
\begin{equation}
\mathbf{\dot{S}}_{\mathbf{k}}^{(1)}=\frac{1}{\hbar}\left[S\mathbf{e}_{z}\times\mathbf{H}_{\mathrm{eff},\mathbf{k}}^{(1)}\right]+\frac{1}{\hbar}\frac{1}{\mathcal{N}}\sum_{i}e^{i\mathbf{k}\cdot\mathbf{r}_{i}}\left[\mathbf{S}_{i}^{(1)}\times\mathbf{H}_{\mathrm{eff},i}^{(0)}\right].
\end{equation}
For small wave vectors, $ka\ll1$ with $a$ being the lattice spacing,
the random occupation numbers average at the wave length of the spin
waves $\lambda=2\pi/k$ and can be replaced by the concentration of
the RE atoms: $p_{i}\Rightarrow\left\langle p_{i}\right\rangle =c$.
Thus one can continue the calculation above and write
\begin{equation}
\mathbf{\dot{S}}_{\mathbf{k}}^{(1)}=\frac{1}{\hbar}\left[S\mathbf{e}_{z}\times\mathbf{H}_{\mathrm{eff},\mathbf{k}}^{(1)}\right]+\frac{1}{\hbar}\left[\mathbf{S}_{\mathbf{k}}^{(1)}\times\mathbf{H}_{\mathrm{eff}}^{(0)}\right],
\end{equation}
where
\begin{equation}
\mathbf{H}_{\mathrm{eff},i}^{(0)}=\left[S\left(J_{0}+D\right)+g\mu_{B}H+c\Sigma J'\right]\mathbf{e}_{z}
\end{equation}
and
\begin{eqnarray}
\mathbf{H}_{\mathrm{eff},\mathbf{k}}^{(1)} & = & \frac{1}{\mathcal{N}}\sum_{i}e^{i\mathbf{k}\cdot\mathbf{r}_{i}}\sum_{j}J_{ij}\sum_{\mathbf{q}}\mathbf{S}_{\mathbf{q}}^{(1)}e^{-i\mathbf{q}\cdot\mathbf{r}_{j}}-cJ'\boldsymbol{\sigma}_{\mathbf{k}}^{(1)}\nonumber \\
 & = & \sum_{\mathbf{q}}\mathbf{S}_{\mathbf{q}}^{(1)}\frac{1}{\mathcal{N}}\sum_{ij}J_{ij}e^{i\mathbf{\left(k-q\right)}\cdot\mathbf{r}_{i}}e^{i\mathbf{q}\cdot\left(\mathbf{r}_{i}-\mathbf{r}_{j}\right)}-cJ'\boldsymbol{\sigma}_{\mathbf{k}}^{(1)}\nonumber \\
 & = & \mathbf{S}_{\mathbf{k}}^{(1)}J_{\mathbf{k}}-cJ'\boldsymbol{\sigma}_{\mathbf{k}}^{(1)},
\end{eqnarray}
where for the square lattice
\begin{equation}
J_{\mathbf{k}}=2J\left[\cos\left(k_{x}a\right)+\cos\left(k_{y}a\right)\right],
\end{equation}
which at small wave vectors simplifies to
\begin{equation}
J_{\mathbf{k}}\cong J_{0}\left[1-\frac{1}{4}\left(ka\right)^{2}\right],\qquad J_{0}=4J.
\end{equation}
Now the equation of motion for $\mathbf{S}_{\mathbf{k}}^{(1)}$ becomes
\begin{eqnarray}
\mathbf{\dot{S}}_{\mathbf{k}}^{(1)} & = & \frac{1}{\hbar}\left[S\mathbf{e}_{z}\times\left(\mathbf{S}_{\mathbf{k}}^{(1)}J_{\mathbf{k}}-cJ'\boldsymbol{\sigma}_{\mathbf{k}}^{(1)}\right)\right]\\
 &  & +\frac{1}{\hbar}\left[\mathbf{S}_{\mathbf{k}}^{(1)}\times\left[S\left(J_{0}+D\right)+g\mu_{B}H+c\Sigma J'\right]\mathbf{e}_{z}\right].\nonumber
\end{eqnarray}
The equation of motion for $\boldsymbol{\sigma}_{\mathbf{k}}^{(1)}$
reads
\begin{eqnarray}
\dot{\boldsymbol{\sigma}}_{\mathbf{k}}^{(1)} & = & \frac{1}{\hbar}\left[-\Sigma\mathbf{e}_{z}\times-J'\mathbf{S}_{\mathbf{k}}^{(1)}\right]\\
 &  & +\frac{1}{\hbar}\left[\boldsymbol{\sigma}_{\mathbf{k}}^{(1)}\times\left(-\Sigma D_{\Sigma}+g'\mu_{B}H-J'S\right)\mathbf{e}_{z}\right].\nonumber
\end{eqnarray}

The next step is to rewrite these equations in terms of the $x$ and
$y$ components:
\begin{eqnarray}
\hbar\dot{S}_{\mathbf{k},x} & = & \varepsilon_{TM}S_{\mathbf{k},y}+ScJ'\sigma_{\mathbf{k},y}\nonumber \\
\hbar\dot{S}_{\mathbf{k},y} & = & -\varepsilon_{TM}S_{\mathbf{k},x}-ScJ'\sigma_{\mathbf{k},x}\nonumber \\
\end{eqnarray}
and
\begin{eqnarray}
\hbar\dot{\sigma}_{\mathbf{k},x} & = & -\Sigma J'S_{\mathbf{k},y}+\varepsilon_{RE}\sigma_{\mathbf{k},y}\nonumber \\
\hbar\dot{\sigma}_{\mathbf{k},y} & = & \Sigma J'S_{\mathbf{k},x}-\varepsilon_{RE}\sigma_{\mathbf{k},x},
\end{eqnarray}
where
\begin{eqnarray}
\varepsilon_{TM} & = & S\left(J_{0}-J_{\mathbf{k}}+D\right)+g\mu_{B}H+c\Sigma J'\\
\varepsilon_{RE} & = & -\Sigma D_{\Sigma}+g'\mu_{B}H-J'S.
\end{eqnarray}
Introducing new variables
\begin{equation}
S_{\pm}\equiv\left(S_{\mathbf{k},x}\pm iS_{\mathbf{k},y}\right),\qquad\sigma_{\pm}\equiv\left(\sigma_{\mathbf{k},x}\pm i\sigma_{\mathbf{k},y}\right),
\end{equation}
one can rewrite the equations in the form
\begin{eqnarray}
\hbar\dot{S}_{+} & = & -i\varepsilon_{TM}S_{+}-iScJ'\sigma_{+}\nonumber \\
\hbar\dot{\sigma}_{+} & = & -i\varepsilon_{RE}\sigma_{+}+i\Sigma J'S_{+},
\end{eqnarray}
as well as the redundant complex-conjugate equations.

Searching for the solution with the harmonic time dependence
\begin{equation}
S_{+}=C_{S}e^{-i\varepsilon t/\hbar},\qquad\sigma_{+}=C_{\sigma}e^{-i\varepsilon t/\hbar},
\end{equation}
one arrives at the system of equations
\begin{eqnarray}
\left(\varepsilon_{TM}-\varepsilon\right)C_{S}+ScJ'C_{\sigma} & = & 0\nonumber \\
-\Sigma J'C_{S}+\left(\varepsilon_{RE}-\varepsilon\right)C_{\sigma} & = & 0.
\end{eqnarray}
Now the energy spectrum is defined by the secular equation (\ref{Secular_equation}).

\section*{\protect \\
}


\begin{thebibliography}{99}
\bibitem{Neel} Louis Néel, Antiferromagnetism and Ferrimagnetism.
Proceedings of the Physical Society, Section A \textbf{65}, 869-885
(1952).

\bibitem{Wolf} W. P. Wolf, Ferrimagnetism, Reports on Progress in
Physics \textbf{24}, 212-303 (1961).

\bibitem{Lectures} E. M. Chudnovsky and J. Tejada, \textit{Lectures
on Magnetism}, Rinton Press (Princeton - NJ, 2006).

\bibitem{Binder-PRB2006} M. Binder,1A. Weber, O. Mosendz, G. Woltersdorf,
M. Izquierdo, I. Neudecker, J. R. Dahn, T. D. Hatchard, J.-U. Thiele,
C. H. Back, and M. R. Scheinfein, Magnetization dynamics of the ferrimagnet
CoGd near the compensation of magnetization and angular momentum,
Physical Review B \textbf{74}, 134404 (2006).

\bibitem{Arena-PRApplied2017} M. E. Jamer, Y. J. Wang, G. M. Stephen,
I. J. McDonald, A. J. Grutter, G. E. Sterbinsky, D. A. Arena, J. A.
Borchers, B. J. Kirby, L. H. Lewis, B. Barbiellini, A. Bansil, and
D. Heiman, Compensated ferrimagnetism in the zero-Moment Heusler alloy
Mn$_{3}$Al, Physical Review Applied \textbf{7}, 064036 (2017).

\bibitem{Siddiqui-PRB2018} S. A. Siddiqui, J. Han, J. T. Finley,
C. A. Ross, and L. Liu, Current-induced domain wall motion in a compensated
ferrimagnet, Physical Review Letters \textbf{121}, 057701 (2018).

\bibitem{Ivanov-review} B. A. Ivanov, Ultrafast spin dynamics and
spintronics for ferrimagnets close to the spin compensation point
(Review), Low Temperature Physics \textbf{45}, 935-962 (2019).

\bibitem{Bonfiglio-PRB2019} G. Bonfiglio, K. Rode, K. Siewerska,
J. Besbas, G. Y. P. Atcheson, P. Stamenov, J. M. D. Coey, A. V. Kimel,
Th. Rasing, and A. Kirilyuk, Magnetization dynamics of the compensated
ferrimagnet Mn$_{2}$Ru$_{x}$Ga, Physical Review B \textbf{100},
104438 (2019).

\bibitem{Kim-Nature2020} C. Kim, S. Lee, H.-G. Kim, J.-Ho. Park,
K.-W. Moon, J. Y. Park, J. M. Yuk, K.-J. Lee, B.-G. Park, S. K. Kim,
K.-J. Kim, and C. Hwang, Distinct handedness of spin wave across the
compensation temperatures of ferrimagnets, Nature Materials \textbf{19},
980-985 (2020).

\bibitem{DavydovaJOP2020} M. D. Davydova, K. A. Zvezdin, A. V. Kimel,
and A. K. Zvezdin, Ultrafast spin dynamics in ferrimagnets with compensation
point, Journal of Physics: Condensed Matter \textbf{32}, 01LT01 (2020).

\bibitem{Yurlov-PRB2021} V. V. Yurlov, K. A. Zvezdin, P. N. Skirdkov,
and A. K. Zvezdin, Domain wall dynamics of ferrimagnets influenced
by spin current near the angular momentum compensation temperature,
Physical Review B \textbf{103}, 134442 (2021).

\bibitem{Chanda-PRB2021} A. Chanda, J. E. Shoup, N. Schulz, D. A.
Arena, and H. Srikanth, Tunable competing magnetic anisotropies and
spin reconfigurations in ferrimagnetic Fe$_{100-x}$Gd$_{x}$ alloy
films, Physical Review B \textbf{104}, 094404 (2021).

\bibitem{Joo-materials2021} S. Joo, R. S. Alemayehu, J.-G. Choi,
B.-G. Park, and G.-M. Choi, Magnetic anisotropy and damping constant
of ferrimagnetic GdCo alloy near compensation point, Materials \textbf{14},
2604 (2021).

\bibitem{Kim-NatMat2022} S. K. Kim, G. S. D. Beach, K.-J. Lee, T.
Ono, T. Rasing, and H. Yang, Ferrimagnetic spintronics, Nature Materials
\textbf{21}, 24-34 (2022).

\bibitem{Guo-PRB2022} M. Guo, H. Zhang, and R. Cheng, Manipulating
ferrimagnets by fields and currents, Physical Review B \textbf{105},
064410 (2022).

\bibitem{Zhang-PRB2022} X. Zhang, B. Cai, J. Ren, Z. Yuan, Z. Xu,
Y. Yang, G. Liang, and Z. Zhu, Spatially nonuniform oscillations in
ferrimagnets based on an atomistic model, Physical Review B \textbf{106},
184419 (2022).

\bibitem{Chen-Materials2025} C. Chen, C. Zheng, S. Hu, J. Zhang,
and Y. Liu, Temperature-dependent compensation points in Gd$_{x}$Fe$_{1-x}$
ferrimagnets, Materials \textbf{18}, 1193 (2025).

\bibitem{Moreno-PRB2025} R. Moreno, P. G. Bercoff, U. Atxitia, R.
F. L. Evans, and O. Chubykalo-Fesenko, Temperature dependence of exchange
stiffness and energy barrier in compensated ferrimagnets, Physical
Review B \textbf{111}, 184416 (2025).

\bibitem{Ciccarelli-AP2025} C. Ciccarelli, G. Nava Antonio, and J.
Barke, Spin emission from antiferromagnets and compensated ferrimagnets,
Applied Physics Reviews \textbf{12}, 041306 (2025).

\bibitem{Bloch} F. Bloch, Zur Theorie des Ferromagnetismus, Zeitschrift
für Physik \textbf{61}, 206-219 (1930).

\bibitem{Griffiths} J. H. E. Griffiths, Anomalous high-frequency
resistance of ferromagnetic metals, Nature \textbf{158}, 670-671 (1946).

\bibitem{Kittel} C. Kittel, Interpretation of anomalous Larmor frequencies
in ferromagnetic resonance experiments, Physical Review \textbf{71},
270-271 (1947).

\bibitem{Walker-PRB1957} L. R. Walker, Magnetostatic modes in ferromagnetic
resonance, Physical Review \textbf{105}, 390-399 (1957).

\bibitem{ABP-SpinWaves} A. I. Akhiezer, V. G. Bar'yakhtar, and S.V.
Peletminskii, {it Spin Waves}, John Wiley \& Sons (1968).

\bibitem{Wangsness-PR1953} R. K. Wangsness, Sublattice effects in
magnetic resonance, Physical Review \textbf{91}, 1085-1091 (1953).

\bibitem{Kittel-PR1959} C. Kittel, Theory of ferromagnetic resonance
in rare earth garnets. III. Giant anisotropy anomalies, Physical Review
B \textbf{117}, 681-687 (1960).

\bibitem{Lin-PRB1988} D. L. Lin and H. Zheng, Spin waves of two-sublattice
Heisenberg ferrimagnets, Physical Review B \textbf{37}, 5394-5400
(1988).

\bibitem{Zhang-JPhys1997} Z.-D. Zhang and T. Zhao, Spin waves at
low temperatures in two-sublattice Heisenberg ferromagnets and ferrimagnets
with different sublattice anisotropies, Journal of Physics: Condensed
Matter \textbf{9}, 8101-8118 (1997).

\bibitem{Karchev-JPhys2008} N. Karchev, Towards the theory of ferrimagnetism,
Journal of Physics: Condensed Matter \textbf{20}, 325219 (2008).

\bibitem{Okuno-APLExpress2019} T. Okuno, S. K. Kim, T. Moriyama,
D.-H. Kim, H. Mizuno, T. Ikebuchi, Y. Hirata, H. Yoshikawa, A. Tsukamoto,
K.-J. Kim, Y. Shiota, K.-J. Lee, and T. Ono, Temperature dependence
of magnetic resonance in ferrimagnetic GdFeCo alloys, Applied Physics
Express \textbf{12}, 093001 (2019).

\bibitem{Haltz-PRB2022} E. Haltz, J. Sampaio, S. Krishnia, L. Berges,
R. Weil, A. Mougin, and A. Thiaville, Quantitative analysis of spin
wave dynamics in ferrimagnets across compensation points, Physical
Review B \textbf{105}, 104414 (2022).

\bibitem{Sanchez-PRB2025} L. Sánchez-Tejerina, D. Osuna Ruiz, V.
Raposo, E. Martínez, L. López Díaz, and Ó. Alejos, Analytical dispersion
relation for forward volume spin waves in ferrimagnets near the angular
momentum compensation condition, Physical Review B \textbf{112}, 104414
(2025).

\bibitem{Pardavi-JMM2000} M. Pardavi-Horvath, Microwave applications
of soft ferrites, Journal of Magnetism and Magnetic Materials \textbf{215-215},
171-183 (2000).

\bibitem{Stanciu-PRB2006} C. D. Stanciu, A. V. Kimel, F. Hansteen,
A. Tsukamoto, A. Itoh, A. Kirilyuk, and Th. Rasing, Ultrafast spin
dynamics across compensation points in ferrimagnetic GdFeCo: The role
of angular momentum compensation, Physical Review B \textbf{73}, 220402(R)
(2006).

\bibitem[(1992)]{Garanin1992} D. A. Garanin, Dynamics of a domain
wall coupled to thermally agitated spins: A model of a rare earth
ferrite-garnet, Zeitschrift für Physik B Condensed Matter \textbf{86},
77-82 (1992).

\bibitem[(1992)]{Garanin1998} D. A. Garanin, Fokker-Planck and Landau-Lifshitz-Bloch
equations for classical ferromagnets, Physical Review B \textbf{55},
3050--3057 (1998).

\bibitem[(2012)]{Chubykalo2012} U. Atxitia, P. Nieves, and O. Chubykalo-Fesenko,
Landau-Lifshitz-Bloch equation for ferrimagnetic materials, Physical
Review B, \textbf{86}, 104414 (2012).

\bibitem[(2023)]{GarChu2023} D. A. Garanin and E. M. Chudnovsky,
Localized spin-wave modes and microwave absorption in random-anisotropy
ferromagnets, Physical Review B \textbf{107}, 134411 (2023).

\bibitem[(2013)]{GarChuPro2013} D. A. Garanin, E. M. Chudnovsky,
and T. Proctor, Random field xy model in three dimensions, Physical
Review B \textbf{88}, 224418 (2013).

\bibitem[(2022)]{GarChu2022} D. A. Garanin and E. M. Chudnovsky,
Random anisotropy magnet at finite temperature, Journal of Physics:
Condensed Matter \textbf{34}, 285801 (2022).

\bibitem{Garanin2025} D. A. Garanin, Energy minima and ordering in
ferromagnets with static randomness, Journal of Physics: Condensed
Matter \textbf{37}, 385803 (2025).

\bibitem{Garanin2021} D. A. Garanin, Energy balance and energy correction
in dynamics of classical spin systems, Physical Review E \textbf{104},
055306 (2021).

\bibitem[(2017)]{Garanin2017} D. A. Garanin, Pulse-noise approach
for classical spin systems, Physical Review E \textbf{95}, 013306
(2017).

\bibitem[(2016)]{Leonov} A. O. Leonov, T. L. Monchesky, N. Romming,
A. Kubetzka, A. N. Bogdanov, and R. Wiesendanger, The properties of
isolated chiral skyrmions in thin magnetic films, New J. Phys.\textbf{
18}, 065003 (2016).

\bibitem[(2023)]{ChuGar2023} E. M. Chudnovsky and D. A. Garanin,
Integral absorption of microwave power by random-anisotropy magnets,
Physical Review B \textbf{107}, 224413 (2023).

\bibitem[(2022)]{Berges2022} Léo Berges. Magnetic skyrmions in GdCo
ferrimagnetic thin-films. Ph. D. Thesis. Université Paris-Saclay,
2022, https://theses.hal.science/tel-03953105v1.

\bibitem[(2021)]{garchu21prb} D. A. Garanin and E. M. Chudnovsky,
Absorption of microwaves by random-anisotropy magnets, Physical Review
B \textbf{103}, 214414 (2021).

\end{thebibliography}
\end{document}